\documentclass[apj]{emulateapj}

\def\tmax{T$_{\rm max}$}
\def\vmax{V$_{\rm max}$}
\slugcomment{Accepted for publication in Astronomical Journal}
\shorttitle{Blazhko effect in RR Lyr stars}
\shortauthors{Le Borgne et al.}
\begin{document}
\title{The all--sky GEOS RR Lyr survey with the TAROT telescopes. \\
  analysis of the Blazhko effect  }
\author{J.-F.~Le Borgne\altaffilmark{1,2,3},
A.~Klotz\altaffilmark{1,2,3,4},
E.~Poretti\altaffilmark{1,2,3,5},
M.~Bo\"er\altaffilmark{6},
N.~Butterworth\altaffilmark{7},
M.~Dumont\altaffilmark{3},
S.~Dvorak\altaffilmark{7},
F.-J.~Hambsch\altaffilmark{3,7,8,9},
F.~Hund\altaffilmark{8},
F.~Kugel\altaffilmark{10},
J.~Vandenbroere\altaffilmark{3},
J.M.~Vilalta\altaffilmark{11}}
\altaffiltext{1}{Universit\'e de Toulouse; UPS-OMP; IRAP; 31400 Toulouse, France}
\altaffiltext{2}{CNRS; IRAP; 14, avenue Edouard Belin, 31400 Toulouse, France}
\altaffiltext{3}{Groupe Europ\'een d'Observations Stellaires (GEOS), 23 Parc de
Levesville, 28300 Bailleau l'Ev\^eque, France}
\altaffiltext{4}{Observatoire de Haute-Provence (CNRS), Saint Michel l'Observatoire, France}
\altaffiltext{5}{INAF-Osservatorio Astronomico di Brera, Via E. Bianchi 46, 23807,
Merate (LC), Italy}
\altaffiltext{6}{ARTEMIS, Universit\'e Nice Sophia-Antipolis, CNRS,
Observatoire de la C\^ote d’Azur, Nice, France}
\altaffiltext{7}{American Association of Variable Star Observers (AAVSO), 49 Bay State
Rd., Cambridge, MA 02138, USA}
\altaffiltext{8}{Bundesdeutsche Arbeitsgemeinschaft f\"ur Veränderliche Sterne e.V. (BAV),
Munsterdamm 90, 12169 Berlin, Germany}
\altaffiltext{9}{Vereniging Voor Sterrenkunde (VVS), Oude Bleken 12, 2400 Mol, Belgium}
\altaffiltext{10}{Observatoire Chante-Perdrix, 04150 Banon, France}
\altaffiltext{11}{Agrupaci\'o Astron\`omica de Sabadell (AAS), Apartat de Correus, 50,
08200 Sabadell (Barcelona), Spain}
\begin{abstract}
We used the GEOS database to study the Blazhko effect of galactic RRab stars.
The database is continuously enriched by maxima supplied
by amateur astronomers and  by a dedicated survey by means of the two
TAROT robotic telescopes.
The same value of the Blazhko period is observed at different values of the pulsation
periods and different values of the Blazhko periods are observed at the same value of the
pulsation period.
There are clues suggesting that the Blazhko effect is changing from one cycle to the next.
The secular changes in the pulsation and Blazhko periods of Z~CVn are anticorrelated.
The diagrams of magnitudes against phases of the maxima
clearly show that the light curves of Blazhko variables can be explained as
modulated signals, both in amplitude and in frequency. The closed
curves describing the Blazhko cycles in such diagrams have different shapes, reflecting the phase shifts
between the epochs of the brightest maximum and the maximum O-C. Our sample shows that
both clockwise and anticlockwise directions  are possible for similar shapes.
The improved observational knowledge of the Blazhko effect, in addition to some
peculiarities of  the light curves, have still to be explained by a satisfactory
physical mechanism.
\end{abstract}
\section{Introduction}
In the  early XX$^{\rm th}$ century S.N.~Blazhko pointed out cycle-to-cycle variations in the light
curves of Var~87.1906 Draconis$\equiv$RW~Dra,
discovered by L.P.~Tserasskaya at Moscow Observatory
\citep{Tsesevich}.
The periodic changes in the times of maximum brightness
were fitted by adding a sinusoidal term to the linear ephemeris, as
happened a few years later when the same effect
was observed in the case of RR~Lyr itself \citep{shapley}.
After being discovered in several other RR Lyr stars, the variations
of the times of maximum brightness (\tmax) and of the magnitudes at maximum
(\vmax) were defined as the Blazhko effect since the list of maxima recorded
from July 15 to August 25, 1906 on RW~Dra
\citep{rwdrab} constituted the first evidence of it.
\citet{klepi} reported on one of the first observational projects to study
the Blazhko effect in a systematic way.
\\
The characteristics of the Blazhko stars discovered in the XX$^{\rm th}$ century
\citep{sze1,sze2,smith} were determined by searching for periodicities
in the \vmax\, and/or O-C ({\it observed} minus {\it calculated} \tmax) values and/or
by performing a frequency analysis of  the photometric  timeseries. In the latter case
the periodic variations of the
shape of the light curve cause a multiplet centered on the
pulsational frequency and with a separation equal to the Blazhko
frequency.
\\
The number of Blazhko stars did not increase significantly
until the advent of CCD techniques. Stars much fainter than
the threshold for photolectric photometry could be
discovered in large-scale surveys as MACHO \citep[MAssive Compact Halo; ][]{macho} and
OGLE \citep[Optical Gravitational Lensing Experiment; ][ and references therein]{ogle3}.
\citet{varsavia} suggested that $\sim$23\% of the RRab stars
in the galactic bulge shows a  Blazhko effect.
\citet{ogle2} substantially confirmed this incidence (27.6\%)
by performing an analysis on a larger sample and a longer baseline.
A much larger incidence was found in the Konkoly Blazhko Surveys (KBSs):
14 out of 30 (47\%) in the KBS~I \citep{budapest} and 45 out of 105 (43\%)
in the KBS~II \citep{granada}. This is probably due to the very small
modulations which could be detected
thanks to the multicolour CCD photometry performed with an automatic
60-cm telescope.
The Konkoly surveys  also provide information on the changes in the
physical parameters during the Blazhko cycle.
\\
The observational scenario changed with the continuous monitoring of
Blazhko RR variables with  the satellites
CoRoT \citep{aql, 793, eli} and {\it Kepler} \citep{itself, szabo}.
The lack of any relevant alias structure in the spectral windows of
the space-based  timeseries allowed us to obtain new results,
as evidence of large cycle-to-cycle
variations of the Blazhko effect \citep{eli}, of the excitation of
additional modes \citep{793}, and of the period doubling bifurcation \citep{szabo}.
\\
Despite these new observational results, the physical explanation of the
Blazhko effect still remain controversial. Very different mechanisms were
invoked: the oblique pulsator model
\citep{don, shiba}, the resonant pulsator  model \citep{dzie},
and the action of a turbulent convective dynamo \citep{sto}. None of these
models is able to provide a close matching between theory and
observations  \citep{geza} and then
it seems that we are still far from a satisfactory theoretical
explanation.
\\
The GEOS ({\it Groupe Europ\'een d'Observations Stellaires})
RR Lyr Database\footnote{The GEOS database is freely accessible
on the internet at the address {\tt http://rr-lyr.ast.obs-mip.fr/dbrr/}} aims
at  collecting the  times of maximum brightness
of RR Lyr stars, exclusively galactic RRab (fundamental radial pulsators)
and RRc (first overtone radial pulsators).
The project was started in 2000 by
archiving recent and historical publications of  \tmax\,
in the literature. It now contains 61000 maxima of
3600 stars  and it is
regularly fed by the valuable inputs from amateur astronomers (BAV, AAVSO, GEOS~...).
We already used the GEOS database to obtain
relevant results in the study of
the evolutionary changes of the pulsational periods of RRab stars \citep{pervar}.
The same data can be used to study
period variations  for a wide sample of galactic RR Lyr stars on a much shorter timescale,
in order to bring out new tiles necessary to  compose the puzzle of the Blazhko effect.
\section{The GEOS RR Lyr survey with TAROT telescopes}
\begin{figure}
\epsscale{.80}
\centerline{   \includegraphics[width=8.5cm]{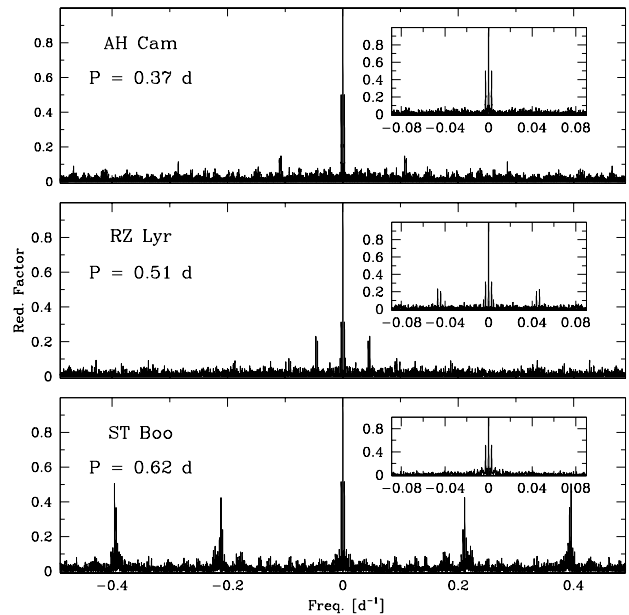} }
\caption{\footnotesize Spectral windows of three RR Lyr stars having different periods.}
\label{sw}
\end{figure}

\begin{figure}
\epsscale{.80}
\centerline{   \includegraphics[width=8.5cm]{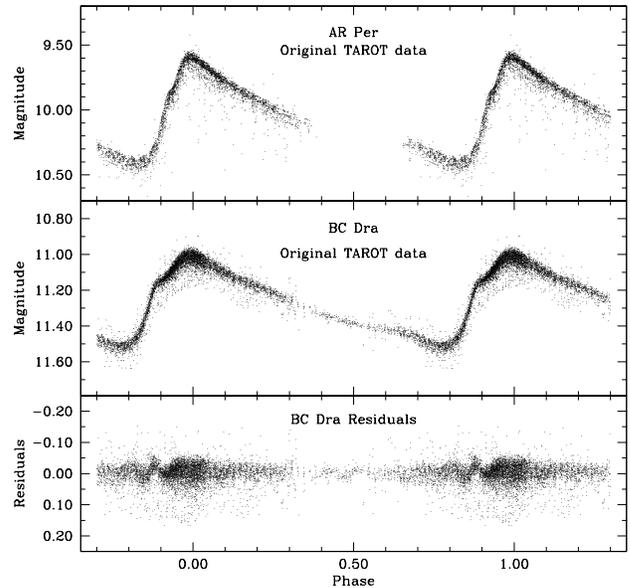} }
\caption{\footnotesize Light curves of the non-Blazhko stars BC Dra
({\it top panel}) and AR Per {(\it middle panel}). The residuals of the
BC Dra data after a  fit with harmonics up to $12f$\, ({\it lower panel})
show the irregular repetitivity of the shoulder on the rising branch.
}
\label{bcdra}
\end{figure}

The main contribution to the GEOS RR~Lyr database comes from the GEOS RR Lyr Survey
and particularly from the observations of the two robotic telescopes TAROT
\citep[{\it T\'elescope \`a Action Rapide pour les Objets
Transitoires};][]{tarot,klotz} located in France (Calern Observatory) and in Chile
(La Silla Observatory). These 25-cm automated telescopes are dedicated to the capture
of the optical afterglows of  Gamma Ray Bursts. Several programs are scheduled
in the idle time, as the one on galactic RR Lyr stars.
Indeed, RR Lyr stars are very suitable targets for small robotic telescopes due to their
short-period and large-amplitude  variability, which allow the observation of several
\tmax\, and \vmax\,  on a single night.
TAROT telescopes have produced about 840 max/year since
the beginning of the observations in 2004. This is about half the total
production of 1640 max/year from ground-based observations on the same period.
This performance appears  more relevant when compared  to an
average  of about 500  max/year over the 110 years elapsed  since the discovery
of RR Lyr stars.
\\
RRab stars have periods in the 0.37-0.65~d range and we could collect from each
TAROT site no more than 1 maximum/night for each given star.
A few equatorial stars could be
observed both from Calern and La Silla, but observing two consecutive maxima is  an
exceptional event due to weather differences.
Since the main goal is the monitoring of as many RR Lyr stars as possible,
the scheduling software
tries to secure points for all the program stars and hence gives  a lesser priority
to a star that has already been observed on a previous night.
The telescopes are continuously moved from one star to the next during the night and
the measurements consist of two consecutive 30-s exposures taken every 10~min.
They are performed around the predicted \tmax\, and the
ephemerides are continuously updated to avoid to miss the real maximum.
The mean error bar of the observed \tmax\, is about 0.003~d (4.3~min).
We obtain an average of 50 measurements in 4.8~hours for each maximum.
\\
The resulting spectral window clearly shows alias peaks at $\pm$1~y$^{-1}=\pm$0.0027~d$^{-1}$ (Fig.~\ref{sw},
inserted boxes), due
to the pointing limitations when stars are too close to the Sun, and different alias structures appear
as a function of the period. The maxima of the shorter-period RR Lyr stars  could
be observed very frequently
and their spectral windows are quite clear, as shown by that of AH Cam (Fig.~\ref{sw},
top panel). Stars with
a period close to 0.50~d could not be observed for a long time when the maxima occur
around sunset or
sunrise and the gaps in the timeseries produced  alias peaks  in their  spectral window
(RZ Lyr, middle panel).
If the period is around or longer than 0.6~d the next observable maximum after
a positive determination occurs a few days later (at least two).
Therefore,
the pseudo-Nyquist frequency is around 0.25~d$^{-1}$, as shown by the case
of ST Boo (lower panel). This extreme case is still largely satisfactory for our purposes,
since the shortest known Blazhko period ($P_B$) is that of SS Cnc, 5.309~d \citep{sscnc}, corresponding to
$f_B$=0.188~d$^{-1}$.
Therefore, we performed our frequency analyses in the interval
0.0--0.2~d$^{-1}$ by means of the
iterative sine--wave least--squares method \citep{vani}.
Due to the scheduling procedure of the TAROT telescopes, photometric
measurements are concentrated around the maxima.
The rising branch of the light curves are  well covered including
the shoulder in some stars currently interpreted as
the effect of  a hypersonic shock-wave propagating in the atmosphere \citep{hyper}.
Our analysis detected such a feature
in the light curves of the non-Blazhko stars BC Dra and AR Per (Fig.~\ref{bcdra}).
BC Dra is a long-period RRab star ($P$=0.72~d), AR Per a short-period one
($P$=0.42~d).
The almost complete phase coverage  of the BC Dra curve allowed us to
determine a pulsational period of 0.71958083$\pm$0.00000015~d.
The amplitudes of the harmonics decrease up to $7f$ and then they show small
fluctuations before the final decline at $12f$.  The residual rms is 0.033~mag and the
residuals  still shows the sudden change in the light curve occurring
on the rising branch (Fig.~\ref{bcdra}, lower panel).
The frequency analysis of the TAROT timeseries and the 105 observed maxima did not
detect any signature of the Blazhko effect.
We applied the same procedure to the AR Per data, thus obtaining
$P$=0.42555060$\pm$0.00000010~d  in the 2004-2011 interval.
This accurate determination is particularly useful since the period of AR Per is subject
to secular changes \citep{pervar}.
As a drawback of the TAROT scheduling procedure,
the phase coverage of the light curves of many program stars is often incomplete, with large
gaps on the descending branch.
\section{The test bench: stars with a known Blazhko effect}
Due to pointing accuracy and small seasonal effects affecting it, the reduction software could
select different stars for the ensemble photometry \citep{damerdji}.
This translates into small systematic differences that in some cases could prevent
the frequency analysis of the \vmax\, values.
The \tmax\, values are free from this effect, since it is
independent from the choice of the reference stars.
However, the O-C values depend on the stability of the main pulsational
period, which could also vary on a time interval as short as seven years.
A linear or parabolic ephemeris could locally fit the \tmax\, values in an unsatisfactory
way \citep{pervar}, leaving some residual peaks at very low frequencies.
In such  cases we corrected this secular  effect
by calculating a linear ephemeris  in the interval covered by the \tmax\, values.
Finally,  we considered both O-C and \vmax\, values
to detect and characterize the Blazhko effect,
giving full confidence to the results only if they
were  in agreement with each other. With the exception of just a couple of cases
which will benefit from a longer time baseline covered by accurate maxima
determinations,
this double-check  ruled out the possibility that the periodicity observed in the O-Cs was caused by
a light-time effect (i.e., a RR Lyr variable in a wide binary),
since \vmax\, values are not expected to show any appreciable variation due
to the orbital motion.  We have a couple of cases where a long-period,  light-time effect
is plausible: TAROT telescopes are monitoring them for a decisive confirmation.
\begin{figure}
\epsscale{.80}
\centerline{   \includegraphics[width=8.5cm]{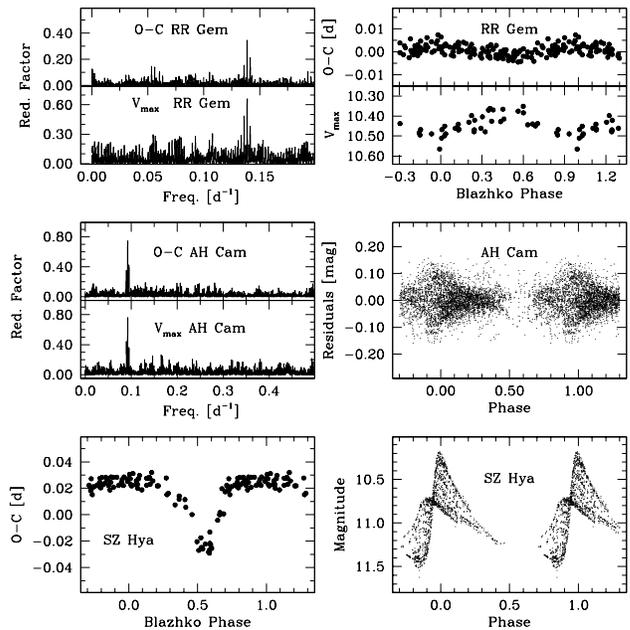} }
\caption{\footnotesize Tools used for the study of the Blazhko effect applied
to the data of RR Gem, AH Cam, and SZ Hya.
}
\label{all}
\end{figure}
\begin{figure}
\epsscale{.80}
\centerline{   \includegraphics[width=8.5cm]{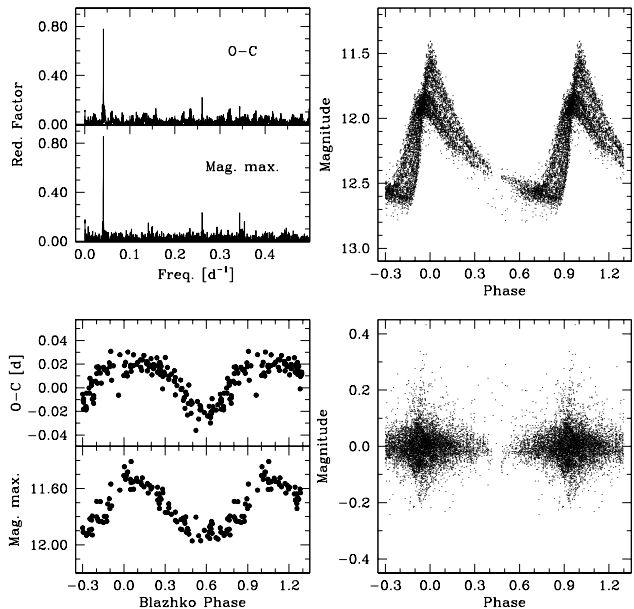} }
\caption{\footnotesize Blazhko effect of BD Dra. {\it Left panels:} power spectra
of O-C and \vmax\, values ({\it top}) and curves of the O-Cs and \vmax\, values ({\it bottom}).
 {\it Right panels:}
TAROT CCD measurements ({\it top})  and residuals after the 24-frequency fit ({\it bottom}).
}
\label{bddra}
\end{figure}
First, we analyzed the stars for which the Blazhko effect was already known.
Figure~\ref{all} gives an overview of the tools applied to some stars
that will be discussed later:
power spectra of  O-C and \vmax\, values (for  RR Gem and AH Cam);
the folded curves of O-C and \vmax\, values (for RR Gem and SZ Hya);
the folded curves of original measurements (for SZ Hya); and
the folded curves of residuals (for AH Cam).
As shown in the upper part of Table~\ref{gen1}, our independent analysis
confirmed most of the previously known results.  For instance, the
very small amplitude Blazhko effect of RR Gem \citep{rrgem} was correctly identified
and characterized (Fig.~\ref{all}, top panels). This not only strengthened
the confidence in our approach, but also  allowed us to build the first
O-C vs. \vmax\, plots for these stars.
Without entering in a detailed discussion of the well-established cases,
we emphasize here some methodological and scientific items.
\subsection{The light curves and the periods of BD Dra and AH Cam} \label{ahc}
The dispersion in the residual plot was also observed on the rising branch
of the Blazhko stars BD Dra
and AH Cam.
Regarding BD Dra, the analysis of 44 old photographic
and visual maxima did not show any significant periodicity. On the other hand, the frequency analysis
of the 136 CCD maxima collected from 2005 to 2011
immediately pointed up  a very clear periodicity of 24.11~d
in both the O-C and \vmax\, values  (left top panel in Fig.~\ref{bddra}).
The O-C and \vmax\,  values folded with the Blazhko period ($P_B$=24.107$\pm$0.001~d)
 show a full amplitude of 0.043~d and 0.40~mag, respectively (left bottom panel in Fig.~\ref{bddra}).
The Blazhko effect is clearly noticeable in the folded light curve
($P_0$=0.58900964$\pm$ 0.00000010~d; right top panel in Fig.~\ref{bddra}).
Very sharp triplets were also detected around $f_0$ and harmonics up to 7$f_0$. It was not possible
to investigate the existence of a quintuplet structure, since such peaks do not stand out
clearly  over the  noise level of 0.003~mag (e.g., the component $f_0+2\,f_B$ has an amplitude
of 0.009 mag only).
The plot of the residuals after a fit with 24 terms ($n\,f_0\pm\,f_B$, with $n$=1,...,~8)
shows a large spread in the values at the phases corresponding to such part of the light curve
(right bottom panel in  Fig.~\ref{bddra}).
\\
The TAROT maxima collected on AH Cam in the recent years clearly indicate that
the true Blazhko frequency is $f_B$=0.092341~d$^{-1}$
(Fig.~\ref{all}, middle panels), i.e., the alias
at +1~y$^{-1}$ of the previous value \citep{pervar}.
This improvement
in the analysis is due to the long coverage ensured by the automatic telescope
during each year, able to strongly damp the aliases at $\pm$1~y$^{-1}$.
The Blazhko frequency was also found in the analysis of the TAROT original
measurements
as triplets centered around $f_0$=2.71211~d$^{-1}$ and  harmonics up to 5$f_0$.
The residuals show a  large scatter in the rising branch (right-middle panel in
Fig.~\ref{all}), although not as large as in the case of BD Dra.
The periods obtained from  the frequency analysis are
$P$=0.36871596$\pm$0.00000006~d and $P_B$=10.8289$\pm$0.0002~d.
\subsection{The O-C and \vmax\, amplitudes of the Blazhko effect}
\begin{figure}
\epsscale{.80}
\centerline{   \includegraphics[width=8.5cm]{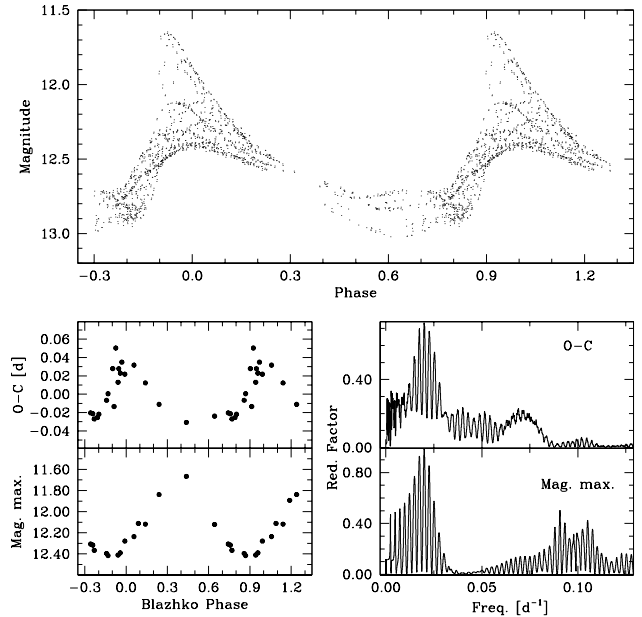} }
\caption{Blazhko effect of XY Eri.
{\it Top panel:} TAROT CCD measurements folded with the pulsational period $P_0$=0.554261~d.
{\it Bottom panels:} power spectra
({\it right}) and folded curves ({\it left})
of O-C and \vmax\, values.
}
\label{xyeri}
\end{figure}
The star with the most spectacular Blazhko effect is XY Eri
(Fig.~\ref{xyeri}). The \vmax\, values
vary  in a range of 0.75~mag, more than half the full amplitude of the
peak-to-peak amplitude, and the O-Cs in a range of 91~min. The Blazhko effect
of IK Hya is very similar, since the O-Cs have the same range, but
the \vmax\, amplitude is probably around 0.5~mag. Since we only observed 9 maxima,
IK Hya deserves further investigation in the future. Most of the stars
shows a \vmax\, amplitude between 0.20 and 0.40~mag
and only RS Boo, SS Oct, and RR Gem have amplitudes smaller than 0.20 mag. These
large variations explain why the Blazhko effect was sometimes referred to
an amplitude modulation \citep[see also the introductory remarks in ][]{sze2}.
Note how the O-C values show variations only in a narrow  phase of the Blazhko
period of SZ Hya (Fig.~\ref{all}, bottom panels). SS CVn  shows a similar variation
of the O-Cs, but its Blazhko period is 3.6 times longer.
\subsection{Long Blazhko periods and long-term modulations} \label{ltime}
\begin{figure}
\epsscale{.80}
\centerline{   \includegraphics[width=8.5cm]{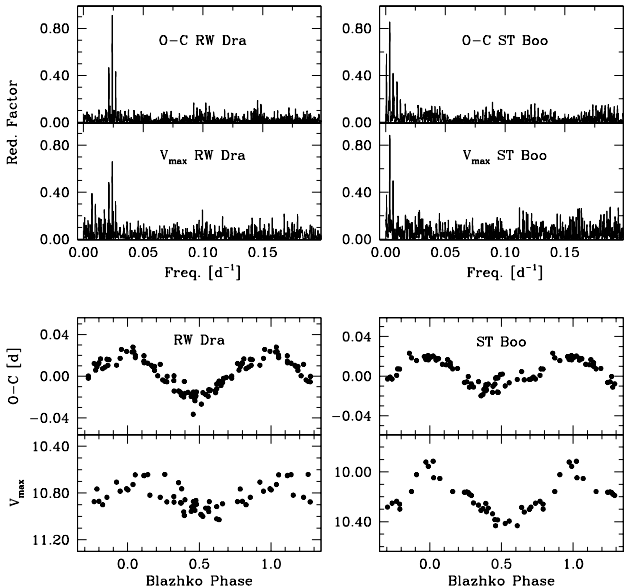} }
\caption{Blazhko effect of RW Dra and ST Boo.
Power spectra and folded curves of the  O-C and \vmax\, values of RW Dra ({\it left panels})
and ST Boo ({\it right panels}).}
\label{rwdrastboo}
\end{figure}

Blazhko periods longer than 100~d  are less common than shorter ones.
In our sample we have the cases of RS Boo, ST Boo, RZ Lyr, RX Cet, and UV Oct.
The longest Blazhko period is that of RS Boo (533~d).
The modern CCD observations provide a spectacular confirmation of the
second longest Blazhko period, that of  ST Boo (Fig.~\ref{rwdrastboo}),
whose features were never presented before.
The analysis of the TAROT CCD timeseries  of RZ Lyr supplies
$P$=0.511241$\pm$0.000005~d and $P_B$=120.19~d, thus confirming
the trends reported by \citet{rzlyr} on a longer time baseline.
In the case of RX Cet, the 15 maxima spread over 7~y did not allow a precise determination of
the Blazhko period, characterized by the second largest amplitude (0.061~d) of  the
O-Cs after that of XY Eri. We detected a tentative period of 273~d superimposed on a slight secular
variation of the period. On the other hand, UV Oct shows a low O-C amplitude of only 0.012~d.
\\
Long, additional  modulations were also claimed as secondary periodicities
superimposed on short Blazhko periods.  For instance,
\citet{gcvs} report on  a composite Blazhko effect for RW Dra.
The O-C values in the GEOS database clearly indicate the Blazhko period of 41.4~d (Fig.~\ref{rwdrastboo}).
The power spectrum and the folded light curve of the residual O-C values are not supporting
the action of other strictly periodic cycles. On the other hand,
the  \vmax\, values show some  scatter, thus suggesting changes in the amplitude from one
cycle to next (Fig.~\ref{rwdrastboo}).
XZ~Dra is perhaps  the extreme case of changes in the Blazhko cycle.
We did not detect any reliable periodicity around 75~d in the  recent maxima collected
in the GEOS database.  The dampening of the Blazhko variation
has been  noticed for more than 50~years in the XX$^{\rm th}$ century \citep{xzdra}.
Since still more  rapid variations are clearly
detected in the continuous space timeseries \citep[e.g., CoRoT 105288363, ][]{eli},
we argue that the composite Blazhko effect and, more  generally, the long-term variations
could be a spurious effect of such cycle-to-cycle changes, irregularly or poorly sampled from ground-based
monitoring.
\subsection{Change in the  Blazhko period with changes in the pulsational period:
Z CVn}
\begin{figure}
\epsscale{.80}
\centerline{   \includegraphics[width=8.5cm]{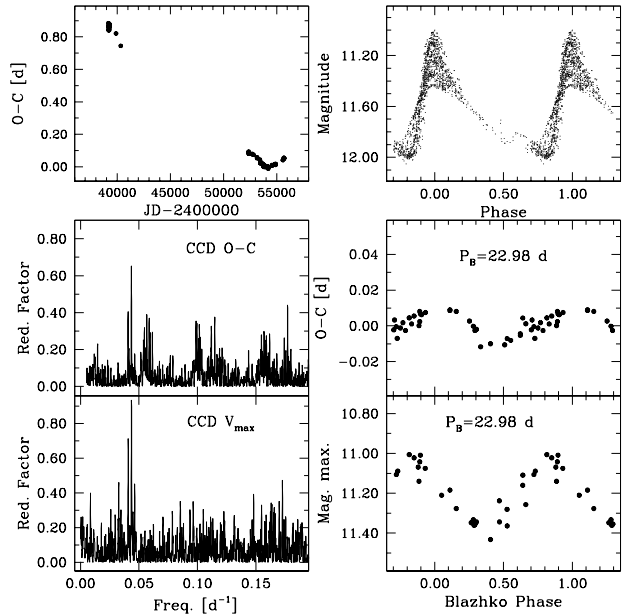} }
\caption{Changes in the Blazhko effect of Z CVn.
{\it Top panel, left:} the plot of the O-Cs with respect to a linear
ephemeris shows a secular change in the period.
{\it Top panel, right:} folded TAROT light curve showing the Blazhko effect.
{\it Bottom panels, left:} the power spectra of the O-C and \vmax\, values
clearly show the peak at $f_B$=0.044~d$^{-1}$.
{\it Bottom panels, right:} O-C and \vmax\, folded with the Blazhko period
corresponding to $f_B$.
}
\label{zcvnbla}
\end{figure}

We detected the Blazhko effect in the CCD data of Z CVn, together with
a well-defined secular change of the pulsation period.
The pulsational period was 0.653942$\pm$0.000007~d around JD 2439000.
The more recent CCD data were obtained in a time interval where the pulsational
period has undergone a sudden change to 0.6538907$\pm$0.0000003~d after JD~2454200
(Fig.~\ref{zcvnbla}, top panel). The Blazhko period around JD~2439000
was 22.75~d \citep{kanyo}. This value is confirmed by our re-analysis of the \tmax\,
values, even if we have to note that the power spectrum shows a broad peak.
When analysing  the CCD data
we immediately noticed that the amplitude of the O-C variation was 0.021~d after JD~2454200,
while it  was 0.033~d in Kanyo's data. This is the first remarkable change in the Blazhko modulation.
More important,
the frequency analysis of the two sets of O-C and \vmax, values supplied a Blazhko  period
of 22.95$\pm$0.03~d after JD~2454200 (Fig.~\ref{zcvnbla}, left-bottom panels). The
amplitude of the O-C values, corrected for the period
change, decreased to 0.021~d, while the amplitude of the CCD \vmax\, values remained
constant at 0.32~mag (Fig.~\ref{zcvnbla}, right-bottom panels).
\\
Our analysis shows that the pulsational period affects  the  Blazhko one.
In particular, the decrease of the pulsational period resulted in a longer Blazhko period, i.e.,
the two changes are anticorrelated. The same behaviour was observed in the cases of
XZ~Cyg \citep{xzcyg},  RV~UMa \citep{rvuma}, RW~Dra \citep{firma}, and  RZ~Lyr \citep{rzlyr}.
However, this does not seem a strict rule, since
correlated changes  were  observed in the case of XZ~Dra \citep{xzdra}.
\begin{figure*}[ht]
\centerline{
   \begin{tabular}{lcr}
   \includegraphics[width=4.9cm]{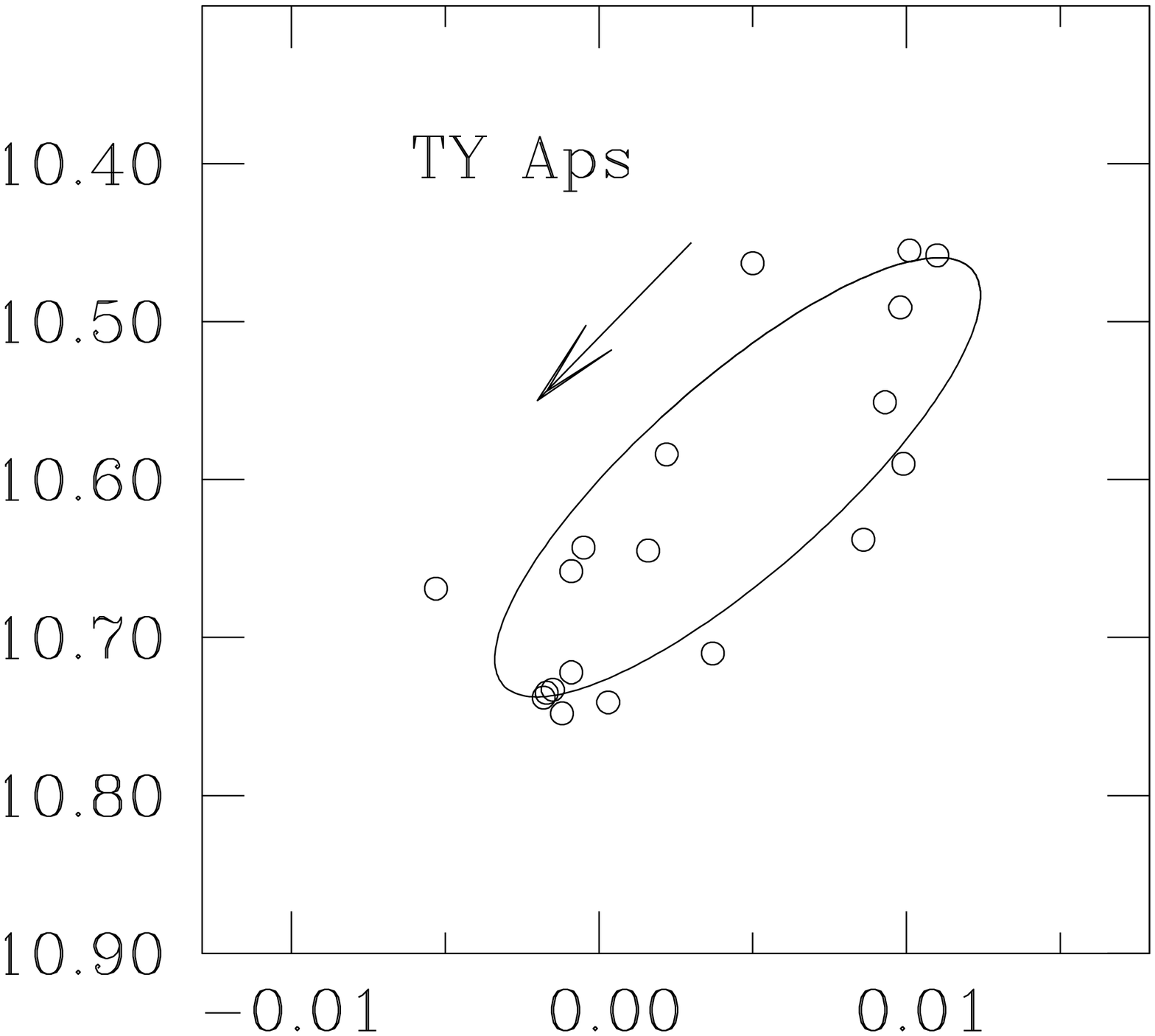}   &
   \includegraphics[width=4.9cm]{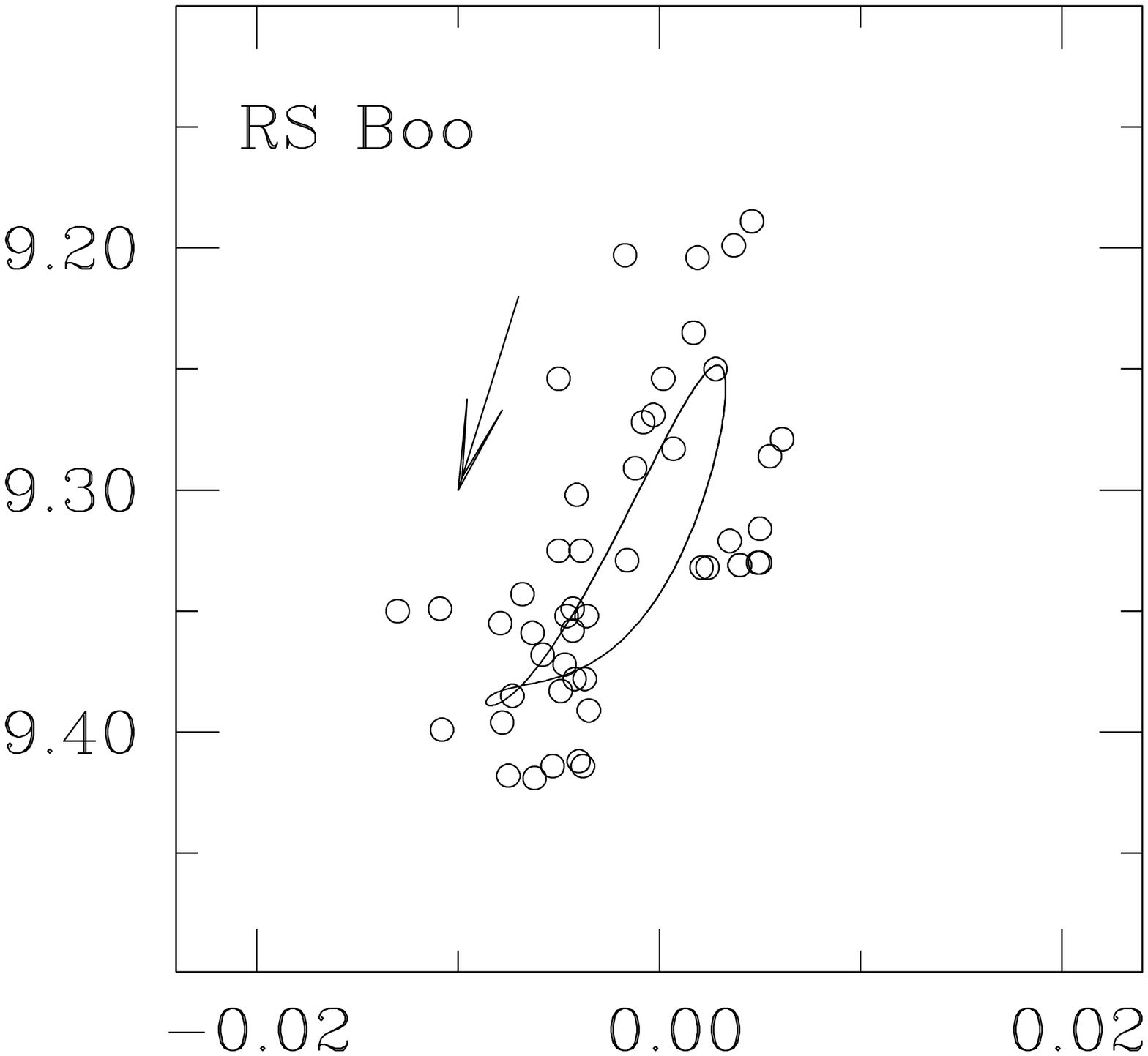}   &
   \includegraphics[width=4.9cm]{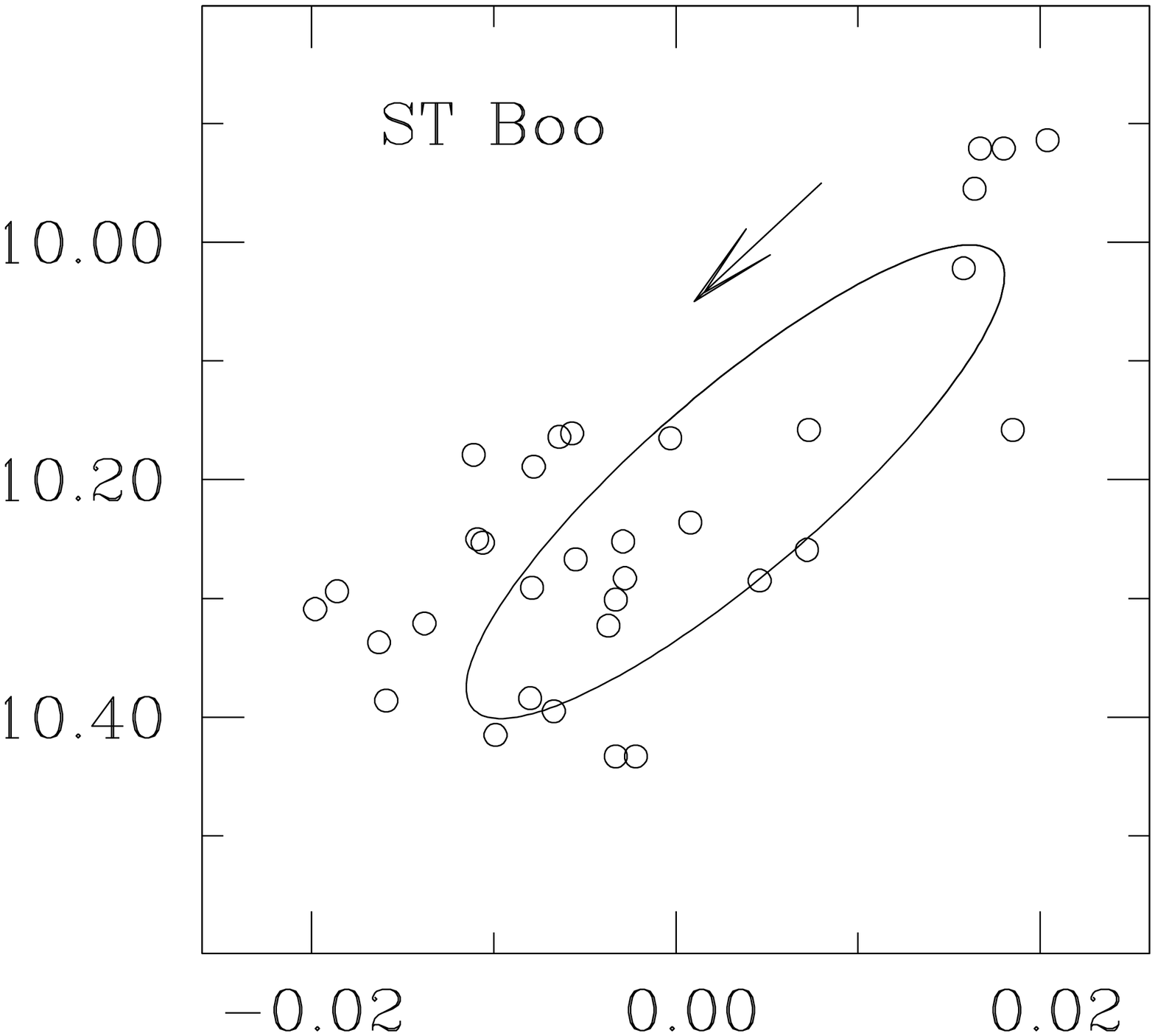}   \\
   \includegraphics[width=4.9cm]{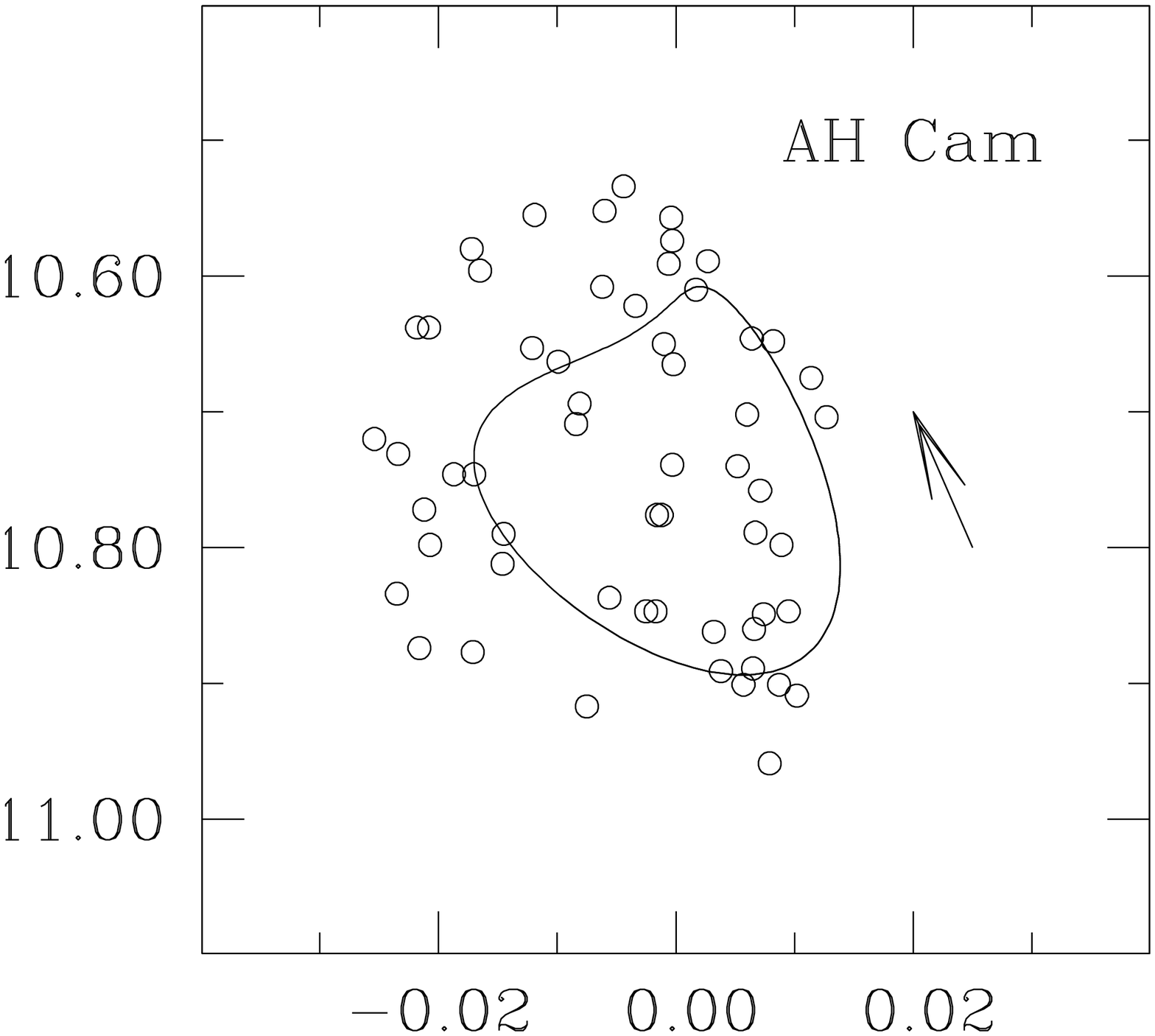}   &
   \includegraphics[width=4.9cm]{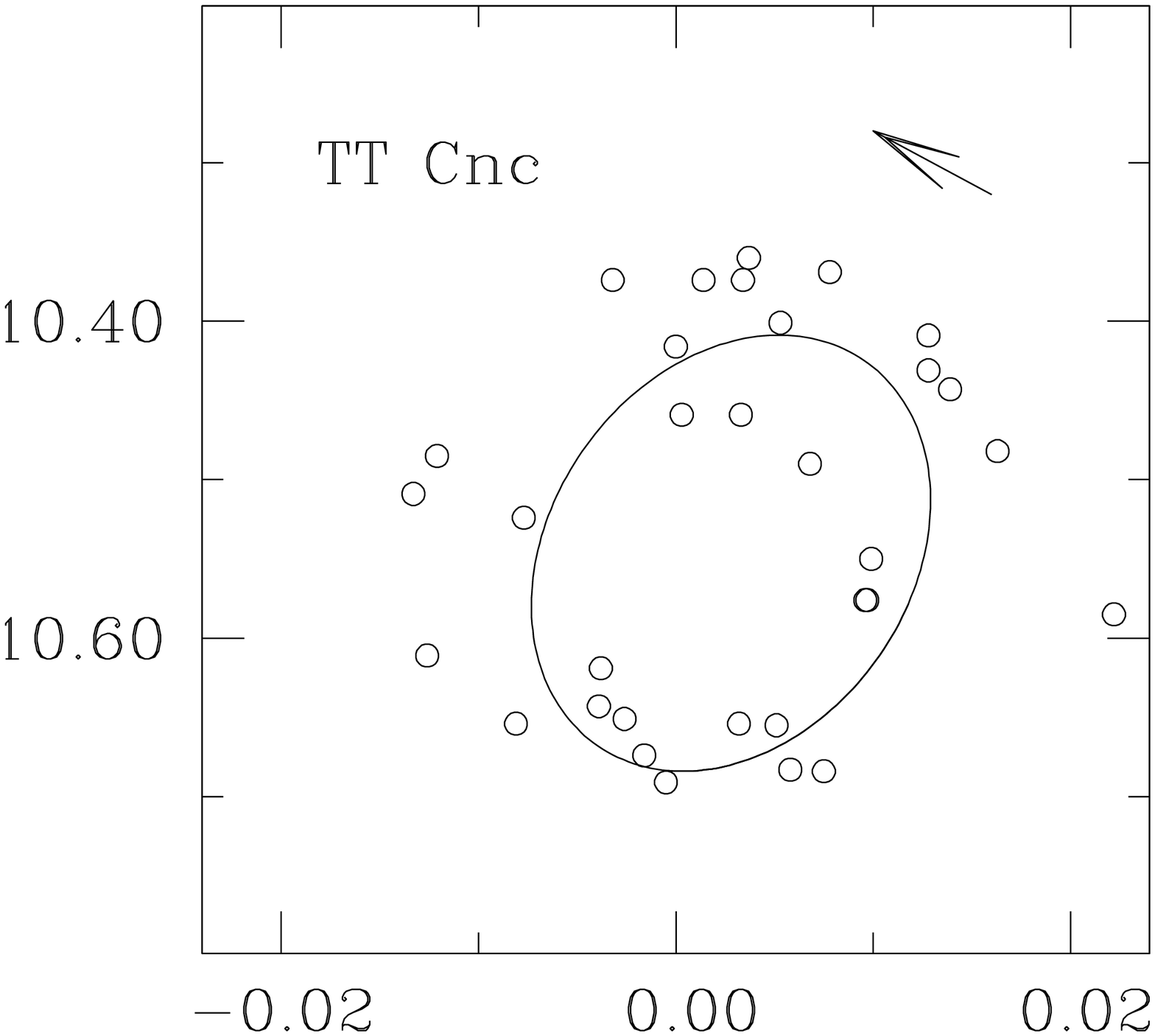}   &
   \includegraphics[width=4.9cm]{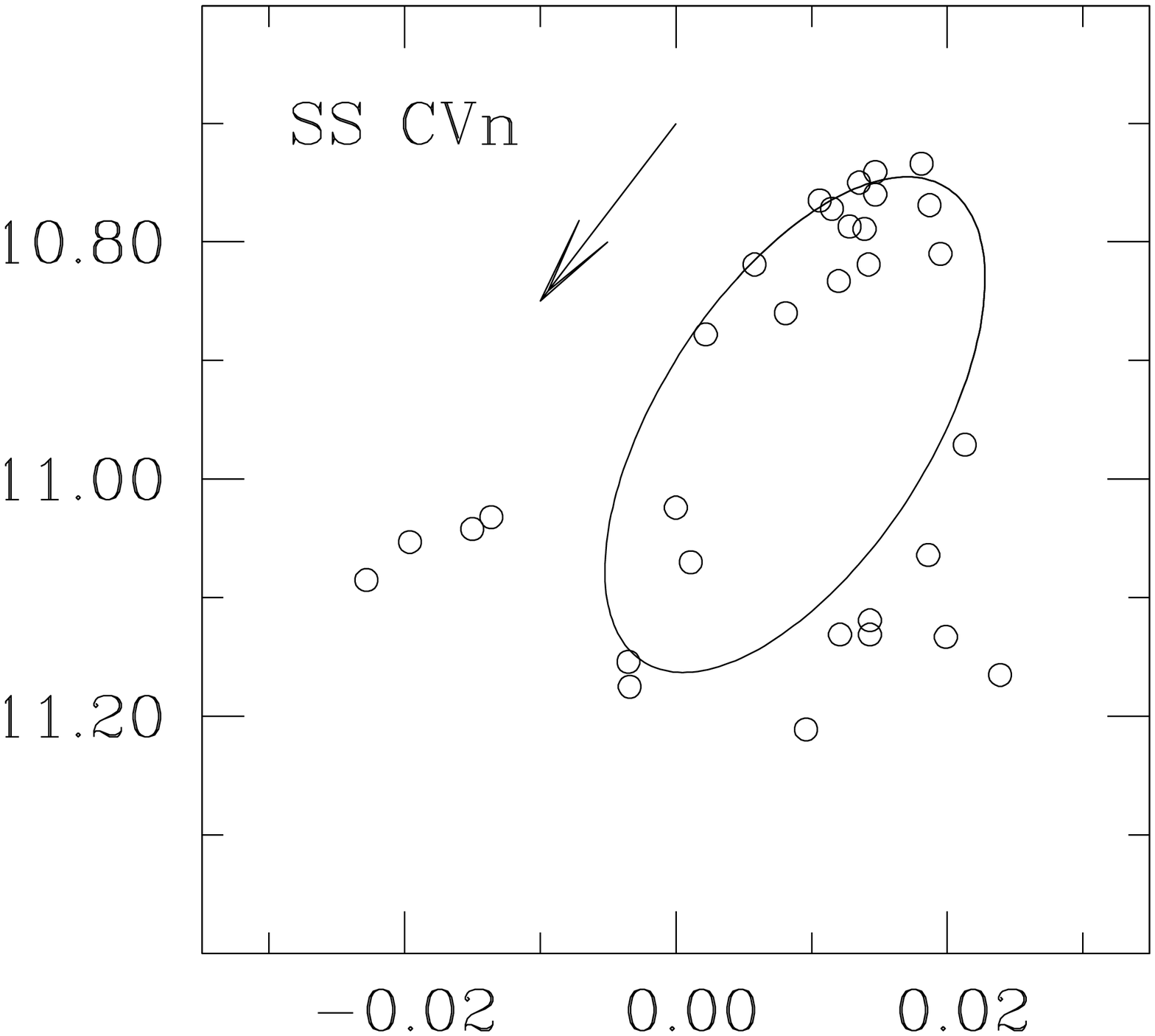}   \\
   \includegraphics[width=4.9cm]{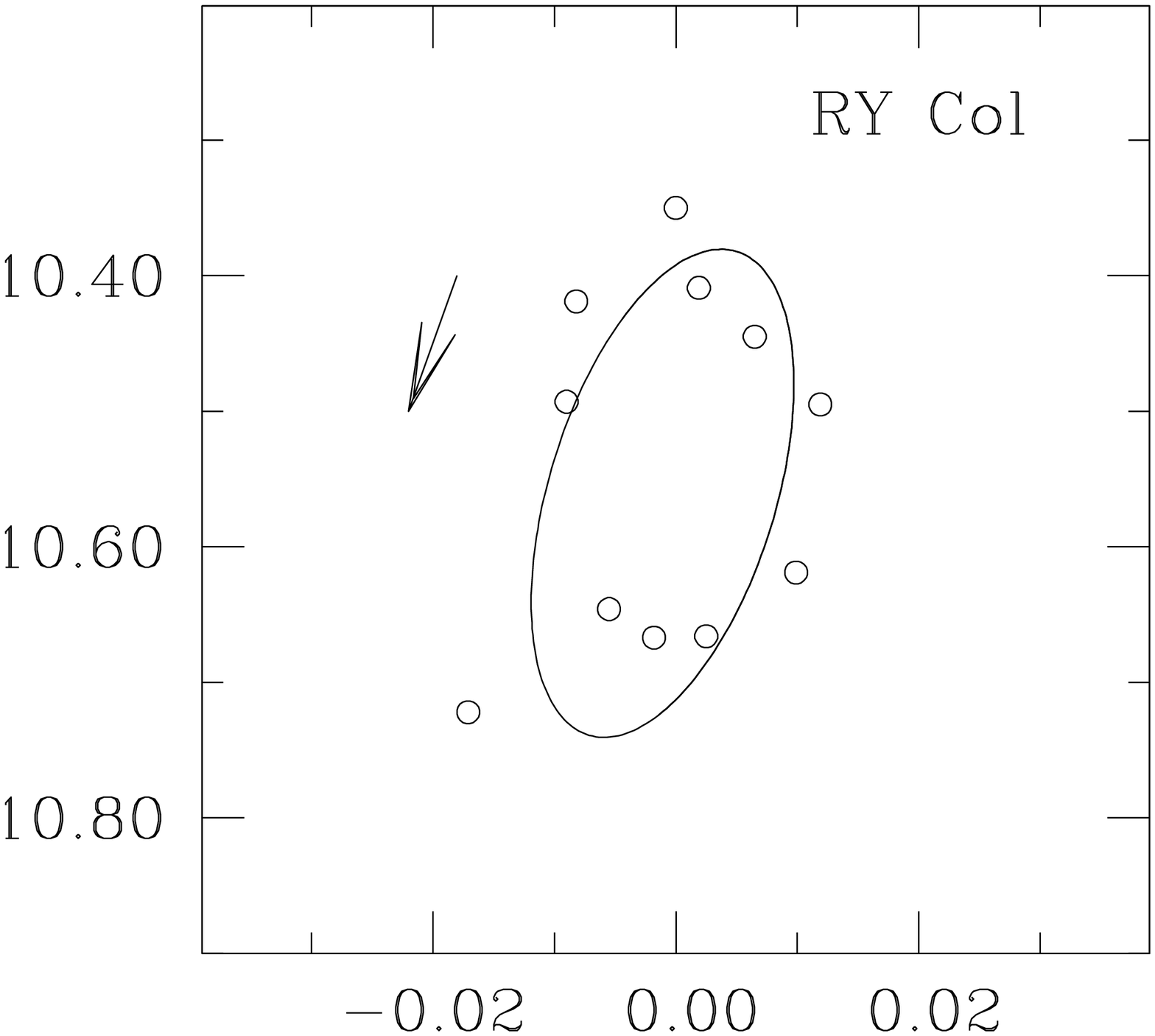}   &
   \includegraphics[width=4.9cm]{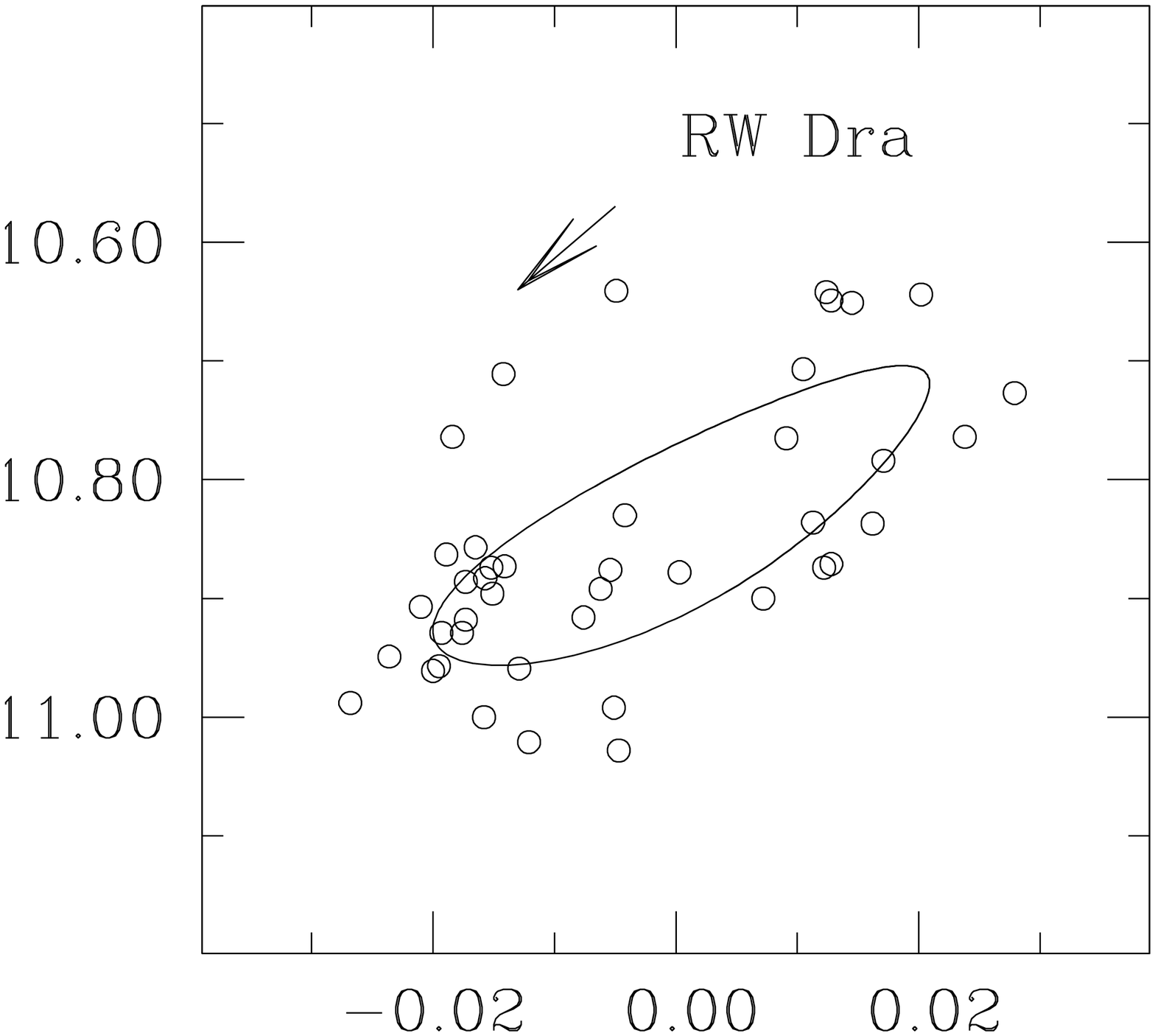}   &
   \includegraphics[width=4.9cm]{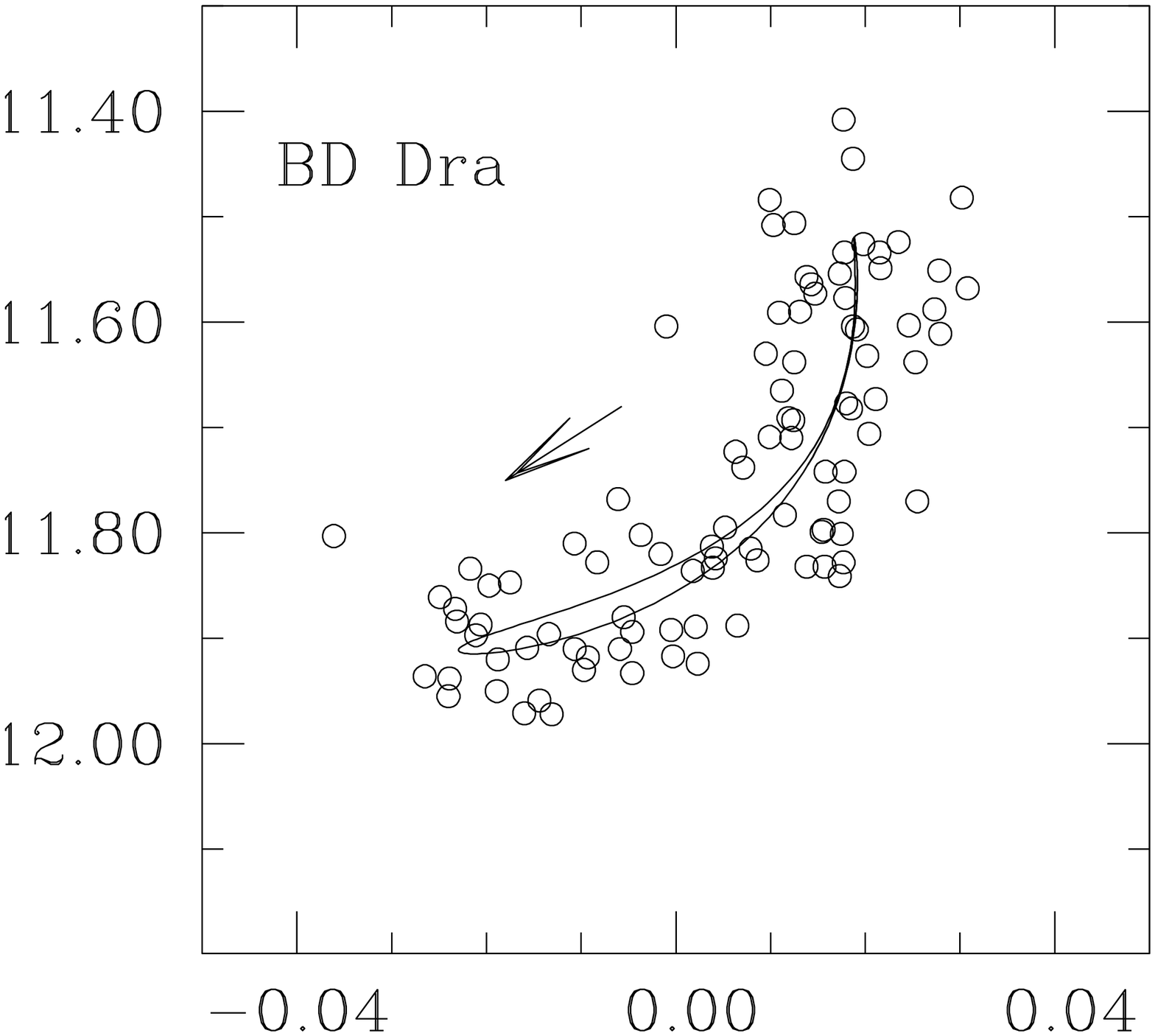}   \\
   \includegraphics[width=4.9cm]{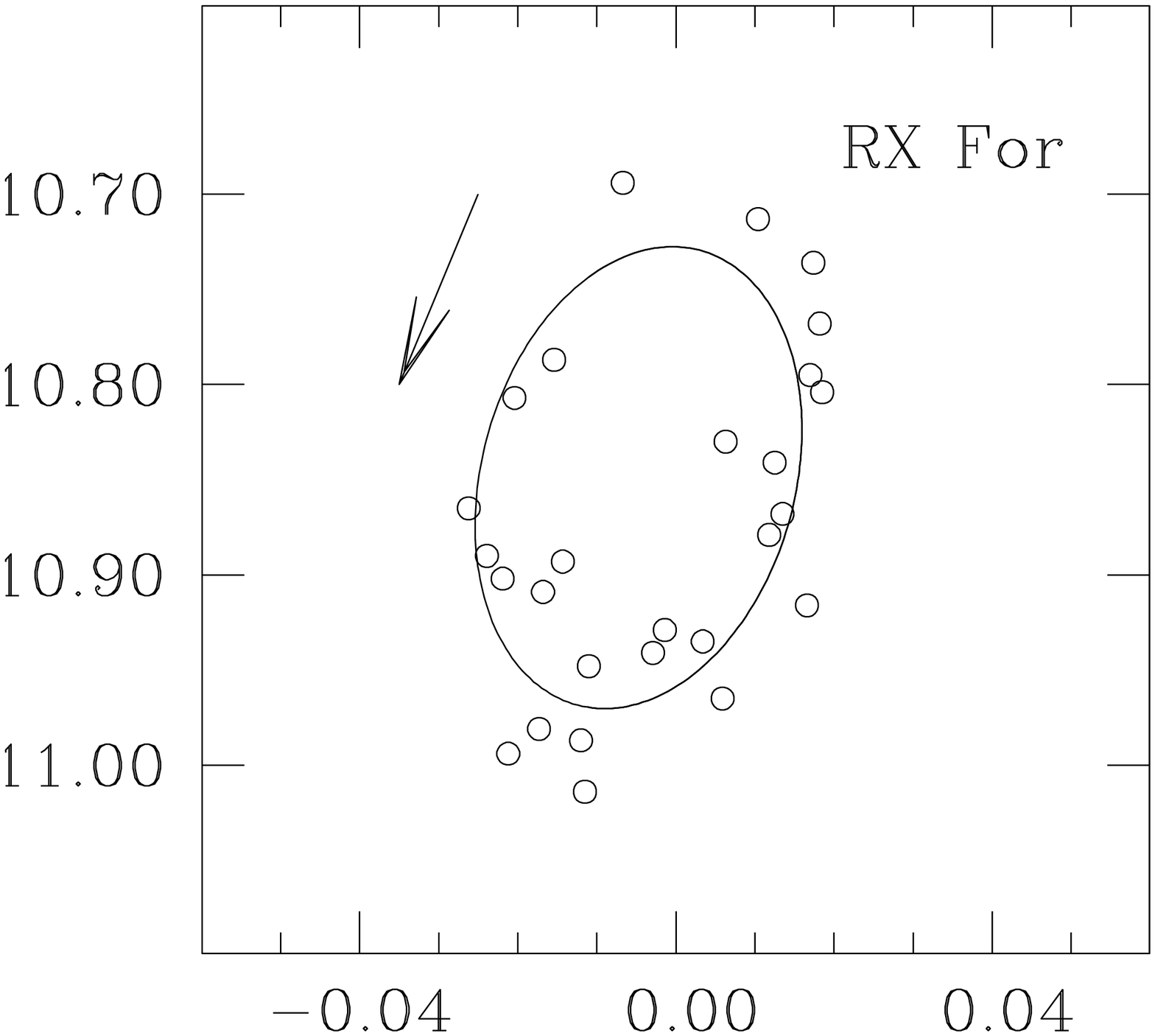}   &
   \includegraphics[width=4.9cm]{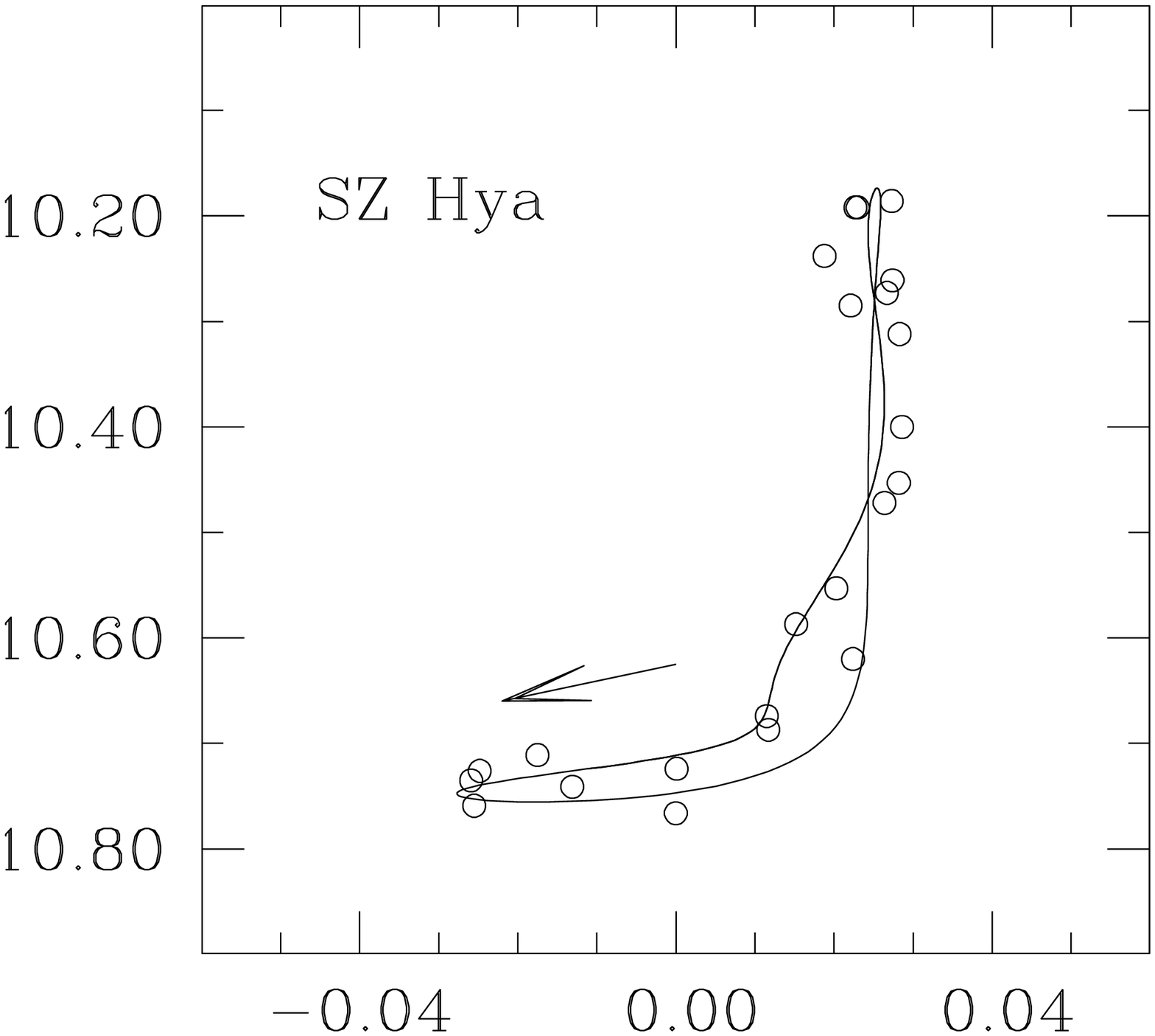}   &
   \includegraphics[width=4.9cm]{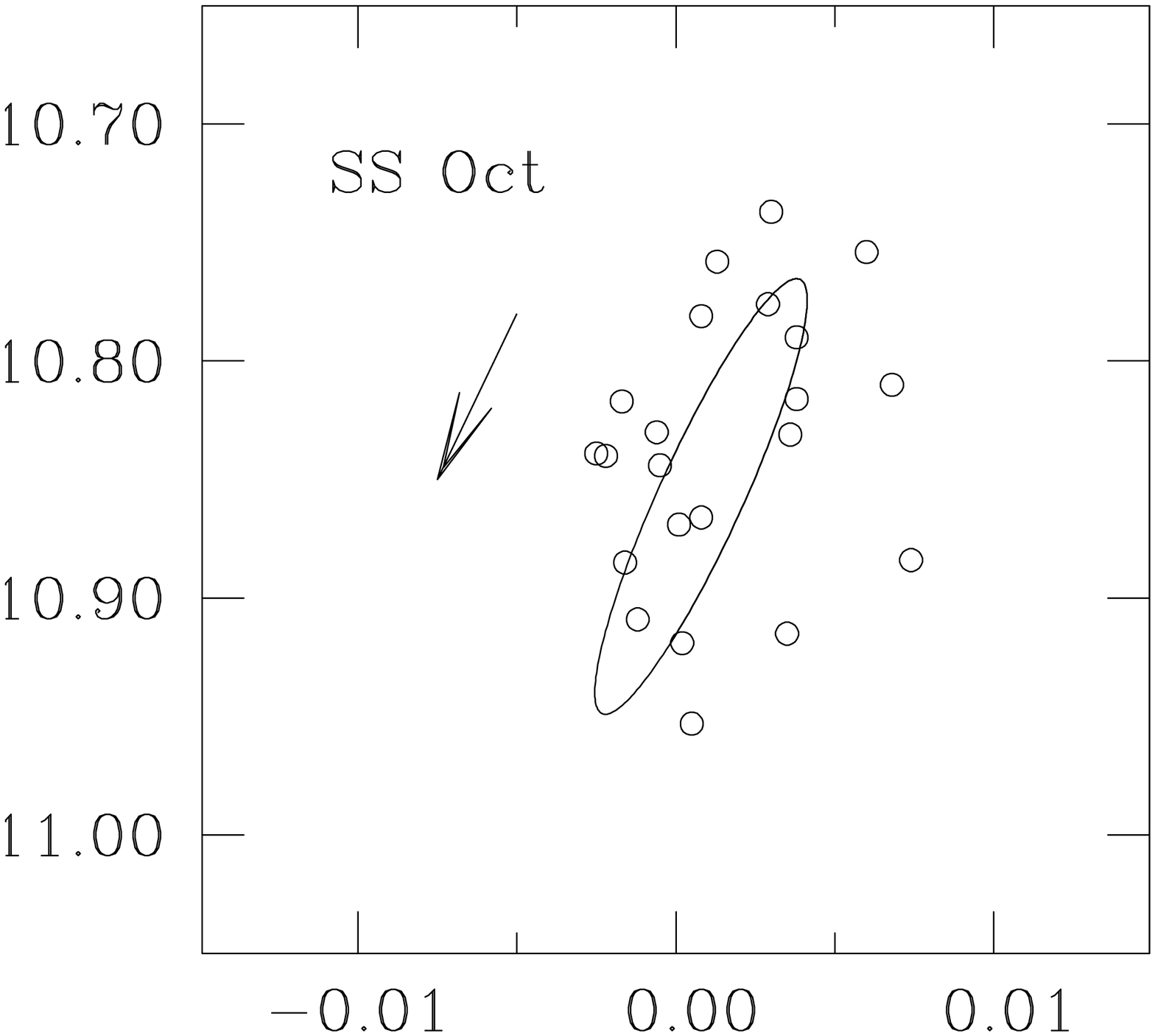}   \\
   &
   \includegraphics[width=4.9cm]{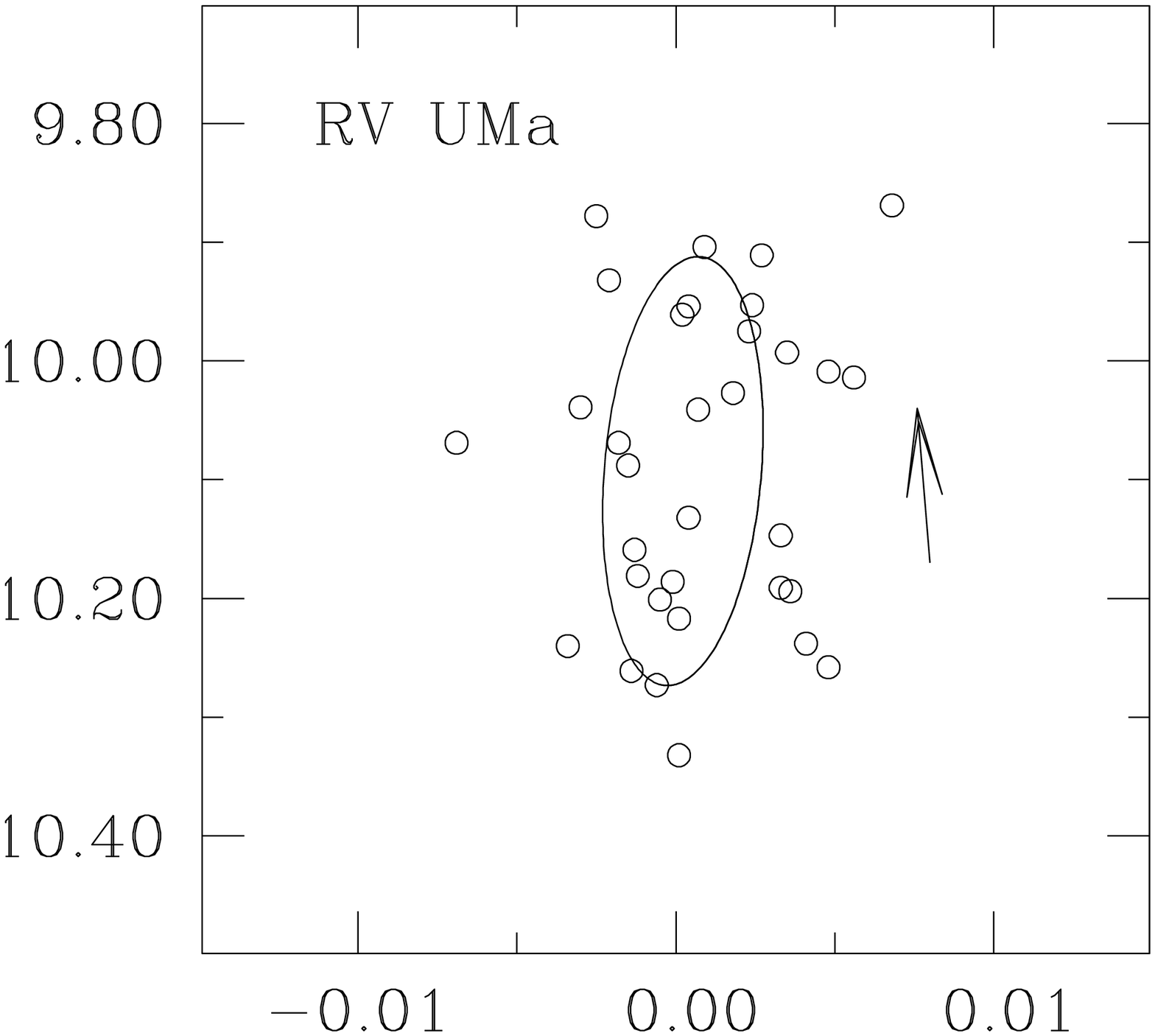}   &
                      \\
   \end{tabular}}
\caption{Stars with the brighest \vmax\, (ordinate)  corresponding to the maximum positive O-C
(abscissa).  The closed curves are run anticlockwise.}
\label{potato1}
\end{figure*}
\section{New Blazhko stars}
Corroborated by the results obtained on stars with a known Blazhko effect,
we applied the same tools to stars poorly observed in the past.
TAROT photometry was able to detect the signatures of the Blazhko effect in
the timeseries of the stars listed in the lower part of Table~\ref{gen1}.
\\
The Blazhko frequencies are determined without ambiguity in the cases of SS CVn, TY Aps,
BI Cen, AF Vel, and SV Vol.
The Blazhko effect of ST Vir is very small, similar to that of RR Gem, and its
detection is another good example of the effectiveness  of the TAROT survey.
In the case of BI Cen, the uncertain value reported by \cite{smith} (70~d) is slightly
different from what we determined. The light curve was well covered by TAROT observations
and we could determine $P$=0.45319439$\pm$0.00000007~d and $P_B$=79.75$\pm$0.02~d
from the frequency analysis of the timeseries. BI Cen shows very sharp  O-C and \vmax\, variations and
the Blazhko variability looks like the most regular in our sample.
\\
The fact that most of the timeseries are single-site introduced some uncertainties
on the determination of  the true Blazhko frequencies, which could be actually
the aliases at  $\pm1$~y$^{-1}$ or $\pm2$~y$^{-1}$ of the highest peak in the
power spectra. This ambiguity results in a real period shorter or longer
by about 10~d for a Blahzko period of 40~d (BS Aps and UU Hya) and 4~d for a period
of 23~d (U Cae).
\begin{deluxetable*}{llrcrccrcrrc}[!t]
\tablecolumns{12}
\tablewidth{0pc}
\tabletypesize{\footnotesize}
\tablecaption{RRab stars showing Blazhko effect observed with TAROTs.}
\tablehead{
\multicolumn{2}{c}{Star} &
\multicolumn{1}{c}{$P_{\rm puls}$} &
\multicolumn{1}{c}{Known $P_B$} &
\multicolumn{1}{c}{N$_{\rm max}$}&
\multicolumn{1}{c}{Ampl.}&
\multicolumn{1}{c}{New $P_B$ }&
\multicolumn{2}{c}{Full Amplitude} &
\multicolumn{2}{c}{Epoch [HJD-2450000]} &
\multicolumn{1}{c}{Phase shift} \\
\multicolumn{2}{c}{} &
\multicolumn{1}{c}{[d]} &
\multicolumn{1}{c}{[d]} &
\colhead{}&
\multicolumn{1}{c}{[mag]}&
\multicolumn{1}{c}{[d]}&
\multicolumn{1}{c}{O-C [d]} &
\multicolumn{1}{c}{Mag. max.} &
\multicolumn{1}{c}{O-C} &
\multicolumn{1}{c}{Mag. max} &
\multicolumn{1}{c}{$\Delta\phi$ [$P_B$]} \\
}
\startdata
\sidehead{Stars with  known Blazhko effect}\\
RS    &  Boo  &  0.3773 & 533$^1$  & 88  & 1.20 & 532.481 & 0.012 &  0.14 & 3519.574 &  3551.523 & -0.06\\
ST    &  Boo  &  0.6222 & 284$^1$  &  58 & 1.35 & 284.090 & 0.032 &  0.40 & 1726.102 &  1748.828 & -0.08\\
AH    &  Cam  &  0.3687 &11.2$^4$ &  88  & 1.06  & 10.829 & 0.031 &  0.29 & 3632.301 &  3635.117 & -0.26\\
TT    &  Cnc  &  0.5634 &89$^1$   & 82   & 0.95 & 89.02   & 0.020 & 0.28    & 2487.945 & 2506.641 & -0.21\\
Z     &  CVn  &  0.6539 & 22.75$^5$&32   & 1.00 & 22.98   & 0.021 &  0.32   & 2745.629 & 2763.781  &  -0.79 \\
Z     &  CVn  &  0.6539 & 22.75$^5$&32   & 1.00 & 22.98   & 0.021 &  0.32   & 2745.629 & 2763.781  &  -0.79 \\
SS    &  CVn  &  0.4785 & 97$^3$  & 39  & 1.25  & 93.72  & 0.028 &  0.42 & 3606.699 &  3620.758  &  -0.15 \\
RV    &  Cet  &  0.6234 & 112$^2$ & 23   &      & 111.732 & 0.034 &  $>$0.20 & 0808.660 & & \\
RX    &  Cet  &  0.5737 & 256$^2$ & 15   &      & 273:    & 0.061 &          & 3145.922 & & \\
RX    &  Col  &  0.5937& 130$^2$      & 20 &  0.85  & 134.77 & 0.032 & 0.29 & 4188.273 &  4228.703  &    -0.30 \\
RY    &  Col  &  0.4788 & 82$^2$  & 33   & 1.20 & 82.080  & 0.022 & 0.36    & 3098.324 & 3112.687 & -0.17\\
VW    &  Dor  &  0.5706 & 24.9$^2$ &  44 & 1.20 & 25.99 &  0.012 &  0.24 & 4095.262 &  4112.027  &    -0.64\\
RW    &  Dra  &  0.4429 & 41.61$^1$&74   & 1.30 & 41.42   & 0.041 & 0.25    & 2371.344 & 2374.656 &   -0.08 \\
BD    &  Dra  &  0.5890 & 24$^3$ & 136 &  1.25  & 24.11 & 0.042  & 0.40 & 3605.965 & 3607.894&   -0.08\\
XY    &  Eri  &  0.5542 & 50$^2$ &  35 &  1.35  & 50.2   & 0.063  & 0.75  & 4143.031 &  4162.109  &    -0.38 \\
RX    &  For  &  0.5973 &31.8$^2$&  35 &  1.20  & 31.790 & 0.041  & 0.24  & 4080.688 &  4087.680  &     -0.22 \\
RR    &  Gem  &  0.3973 &7.230$^6$& 132&  1.25  &  7.209 & 0.004  & 0.11  & 1936.137 &  1939.957  &      -0.53\\
SZ    &  Hya  &  0.5372 &  26.3$^2$  &  85 & 1.35 &  26.233 & 0.053 & 0.58 & 2054.242 &  2054.898& -0.02 \\
IK    &  Hya  &  0.6503 &  72$^2$    &   9 & 0.80 &  67.5:  & 0.063 & $\geq$0.45     & 4319.551 &            &  \\
RZ    &  Lyr  &  0.5112 & 116.7$^1$ & 101 & 1.25 & 120.190 & 0.017 & 0.44 & 0072.211 &  0152.137& -0.66\\
SS    &  Oct  &  0.6219 & 145$^2$   & 28  &  1.00 & 144.93&  0.007 & 0.18 & 4162.660 &  4172.805  &    -0.07 \\
UV    &  Oct  &  0.5425 & 145$^2$   & 30 &   1.10 & 146.99&  0.012 & 0.39 & 5548.223 &  5609.961  &    -0.42 \\
RV    &  UMa  &  0.4681 & 90.63$^7$&  87   & 1.30 & 90.170  & 0.005 &  0.36 & 0955.316 &  0975.152  &    -0.22 \\
AF    &  Vel  &  0.5274&  59$^2$  & 30 &  1.15  &  58.548& 0.017 & 0.30 &  4181.438 &   4212.176 &     -0.52\\
\sidehead{New Blazhko RR Lyr stars}
TY & Aps &  0.5017&            & 25 &  1.15  & 109.13 & 0.016 & 0.28 &  4340.516 &  4350.883  &    -0.09 \\
BS & Aps &  0.5826&            & 24 &  0.80  &  40.93 & 0.021 &  0.24 &3166.379 &  3182.137  &    -0.38 \\
U  & Cae &  0.4198&            & 33 &  1.35  &  22.80 & 0.007 & 0.25 & 4459.238 &  4473.602   &   -0.63 \\
BI & Cen &  0.4532&            & 35 &  1.25  &  79.68 & 0.021 & 0.30 &  4663.672 &   4695.944 &     -0.40 \\
UU & Hya &  0.5239&            & 30 &  1.20  &  39.89 & 0.028 & 0.26 &  2367.930 &  2401.035  &    -0.83 \\
DD & Hya &  0.5018&            & 25 &  1.30  &  34.59 & 0.016 & 0.33 & 2829.426 &  2847.238  &    -0.52 \\
RY & Oct &  0.5635             &    & 18     &   1.25 & 216.45  & 0.019 & & 4384.359 \\
ST & Vir &  0.4108&            & 28 &  1.30  &  25.58  & 0.008  & 0.10& 3484.652 &  3495.016  &    -0.40 \\
SV & Vol &  0.6099&            & 31 &  0.90  &  85.47 & 0.009 & 0.18 & 4230.387 &  4282.523  &    -0.61 \\
\enddata
\tablecomments{
References: $^1$\citet{gcvs}; $^2$\citet{sodor}, $^3$\citet{lloyd},
$^4$\citet{pervar}, $^5$\citet{kanyo}, $^6$\citet{rrgem}, $^7$\citet{rvuma}}
\label{gen1}
\end{deluxetable*}
\begin{figure}[ht]
\centerline{
   \begin{tabular}{lr}
   \includegraphics[width=4cm]{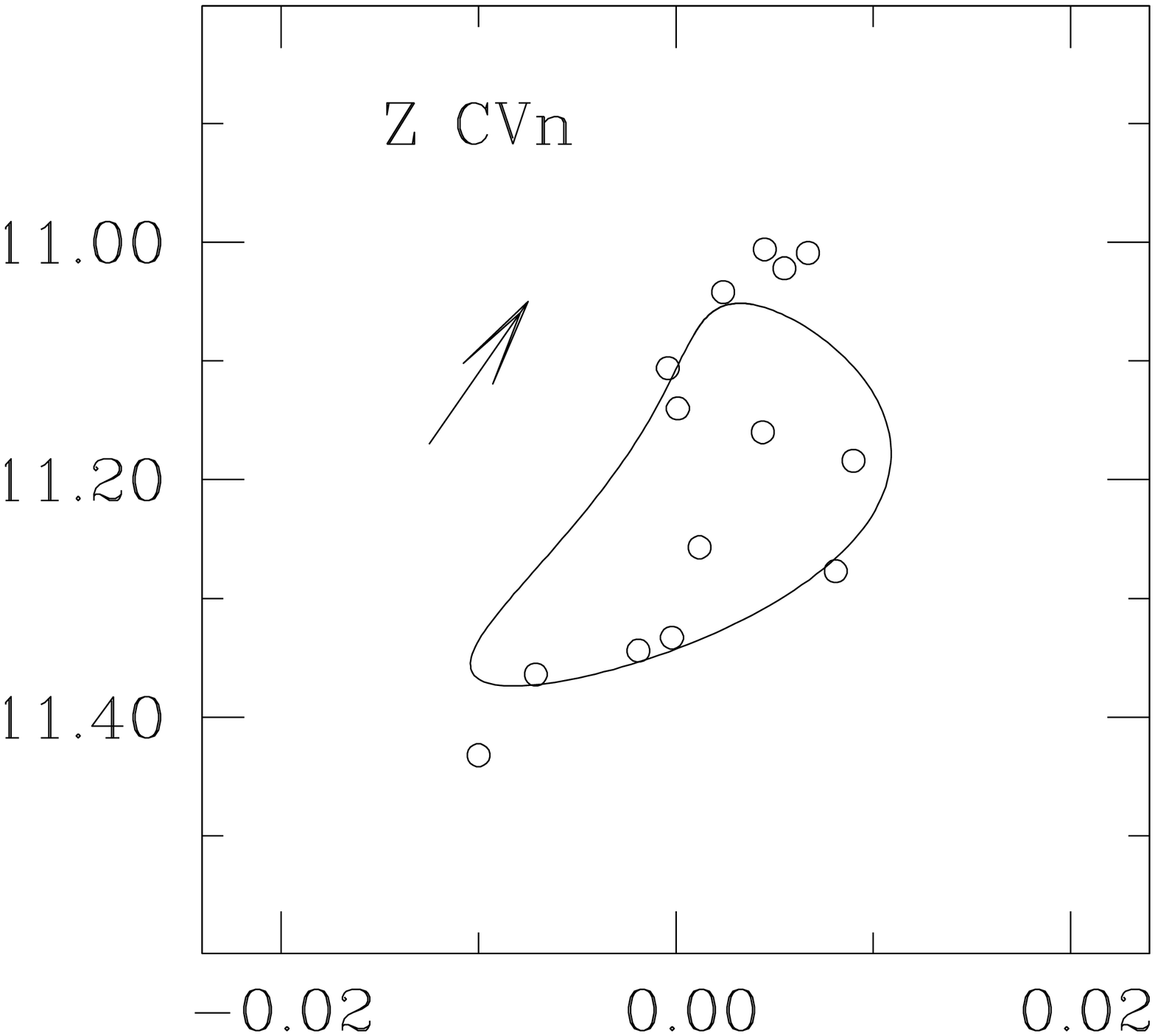}  &
   \includegraphics[width=4cm]{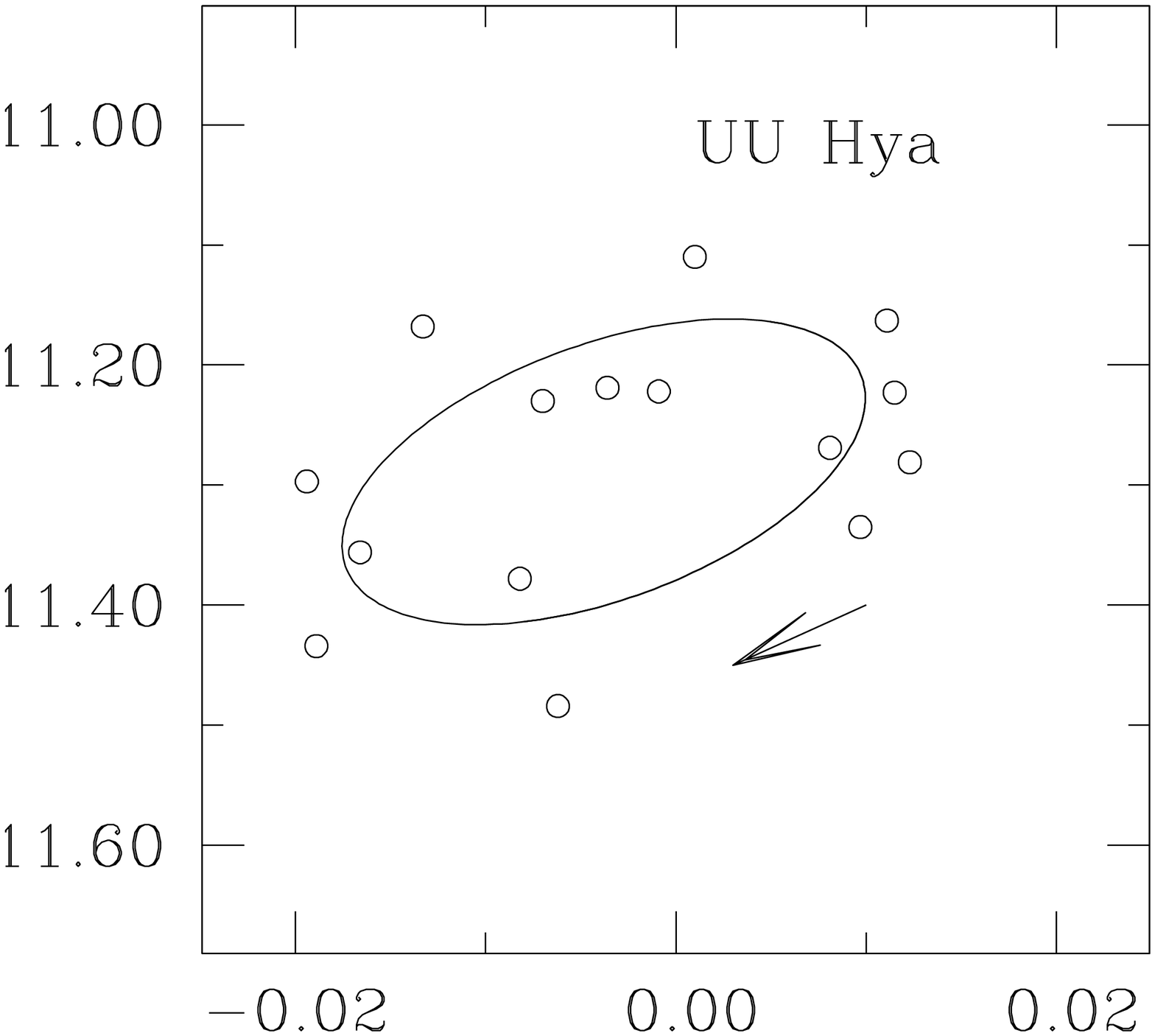}  \\
   \end{tabular}
}
\caption{Same as Fig.~\ref{potato1}, but the closed curves are run in the clockwise direction.}
\label{potato2}
\end{figure}
\begin{figure*}[ht]
\centerline{   \begin{tabular}{lcr}
   \includegraphics[width=4.9cm]{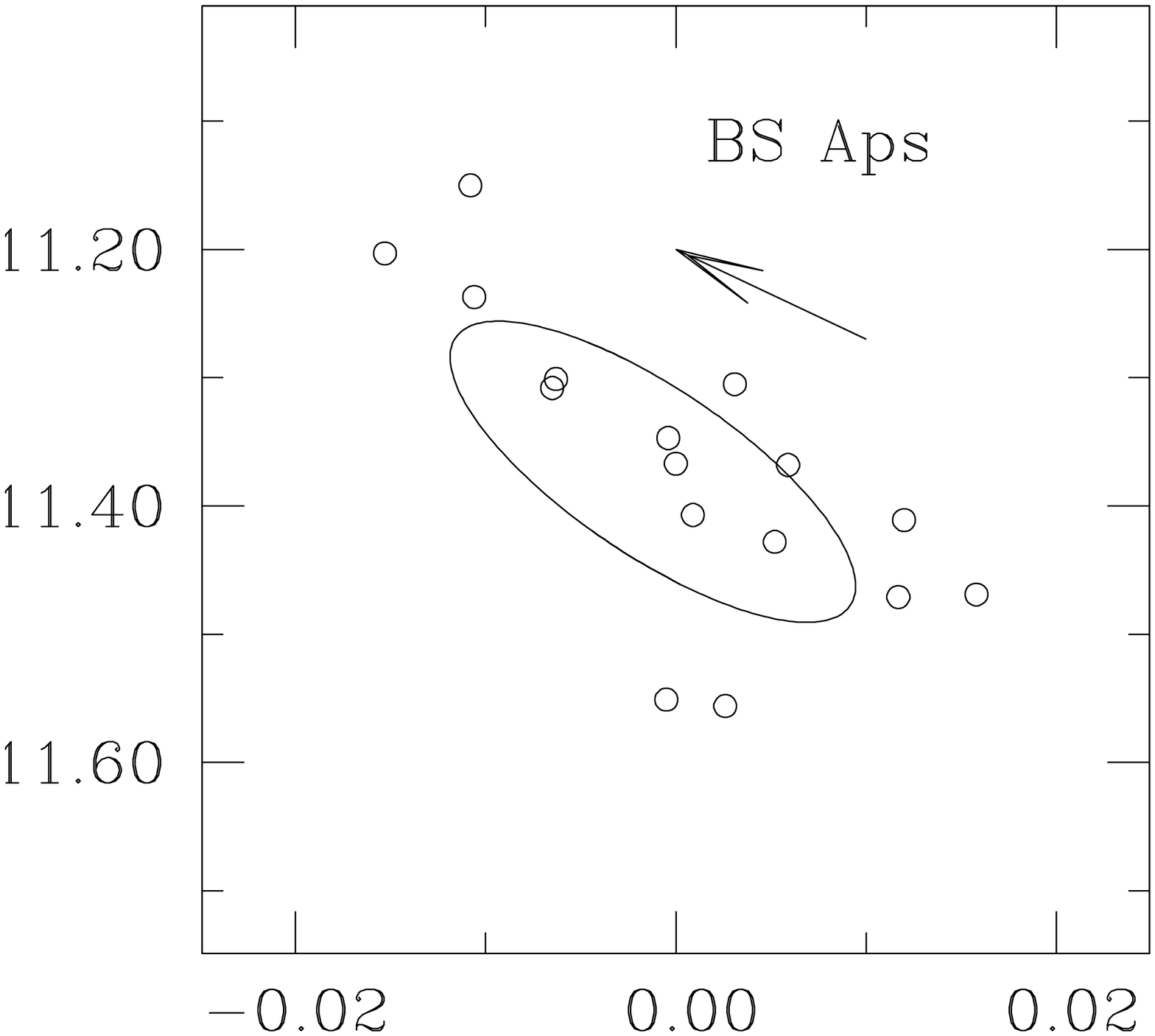}  &
   \includegraphics[width=4.9cm]{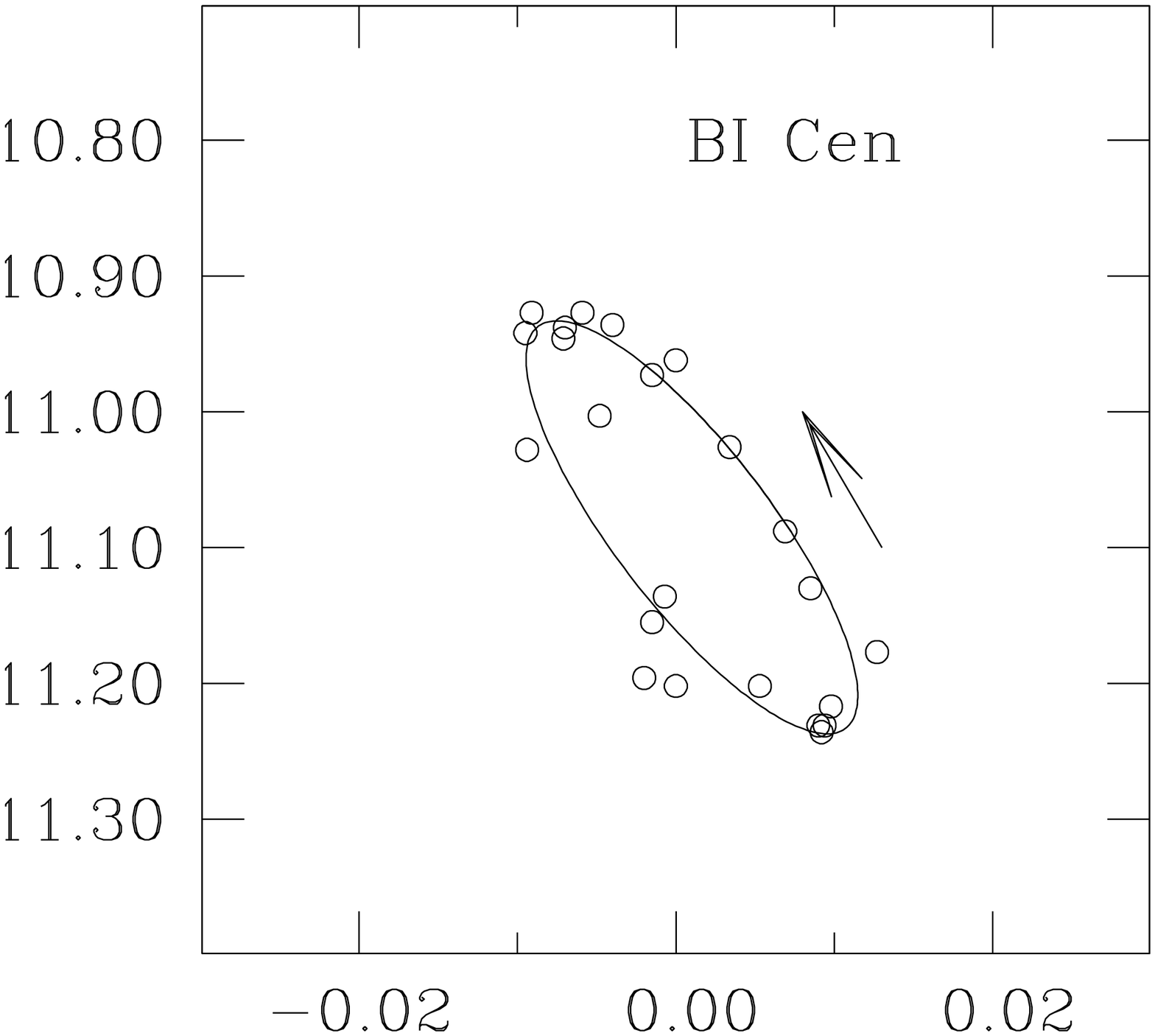}  &
   \includegraphics[width=4.9cm]{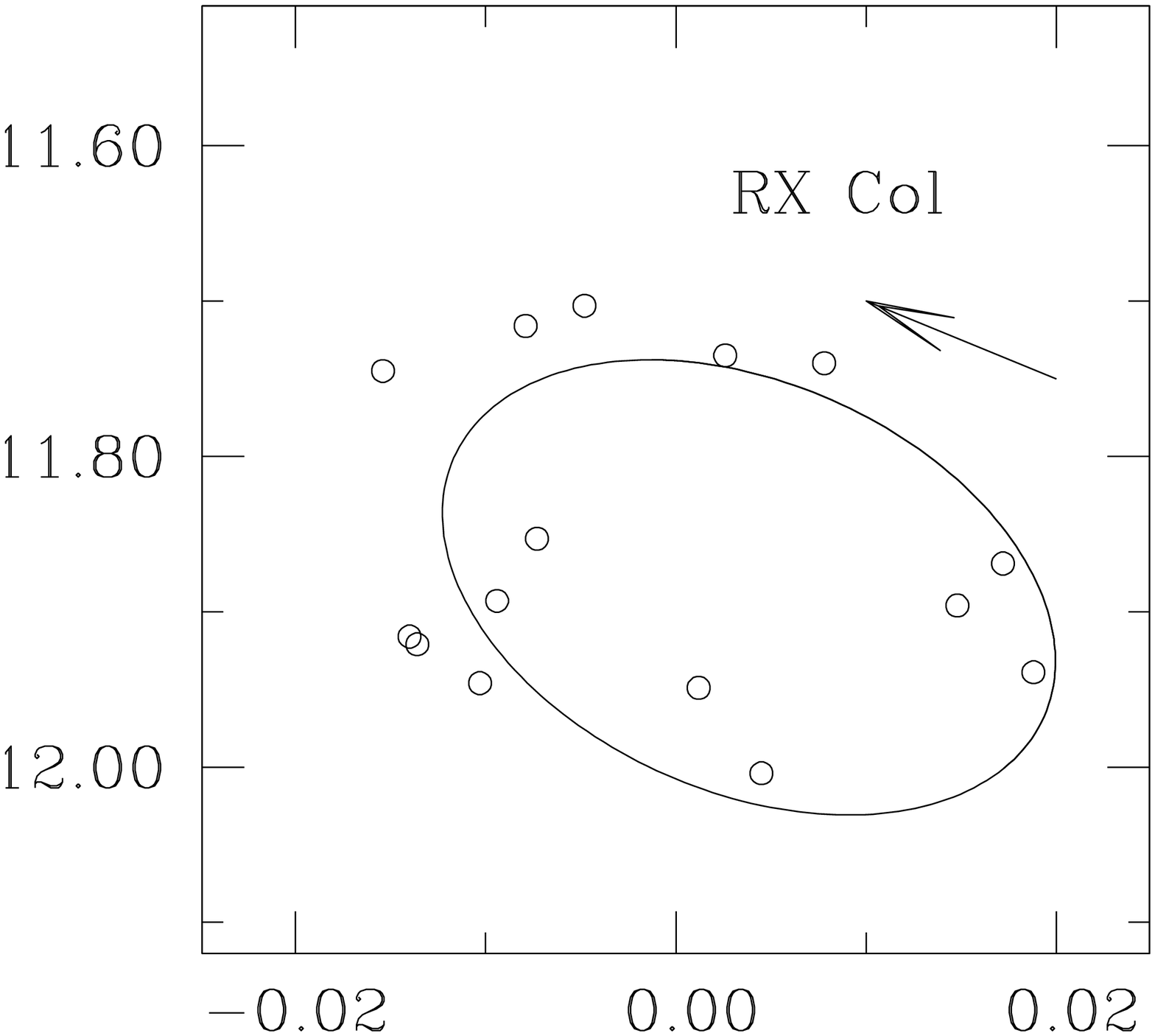}  \\
   \includegraphics[width=4.9cm]{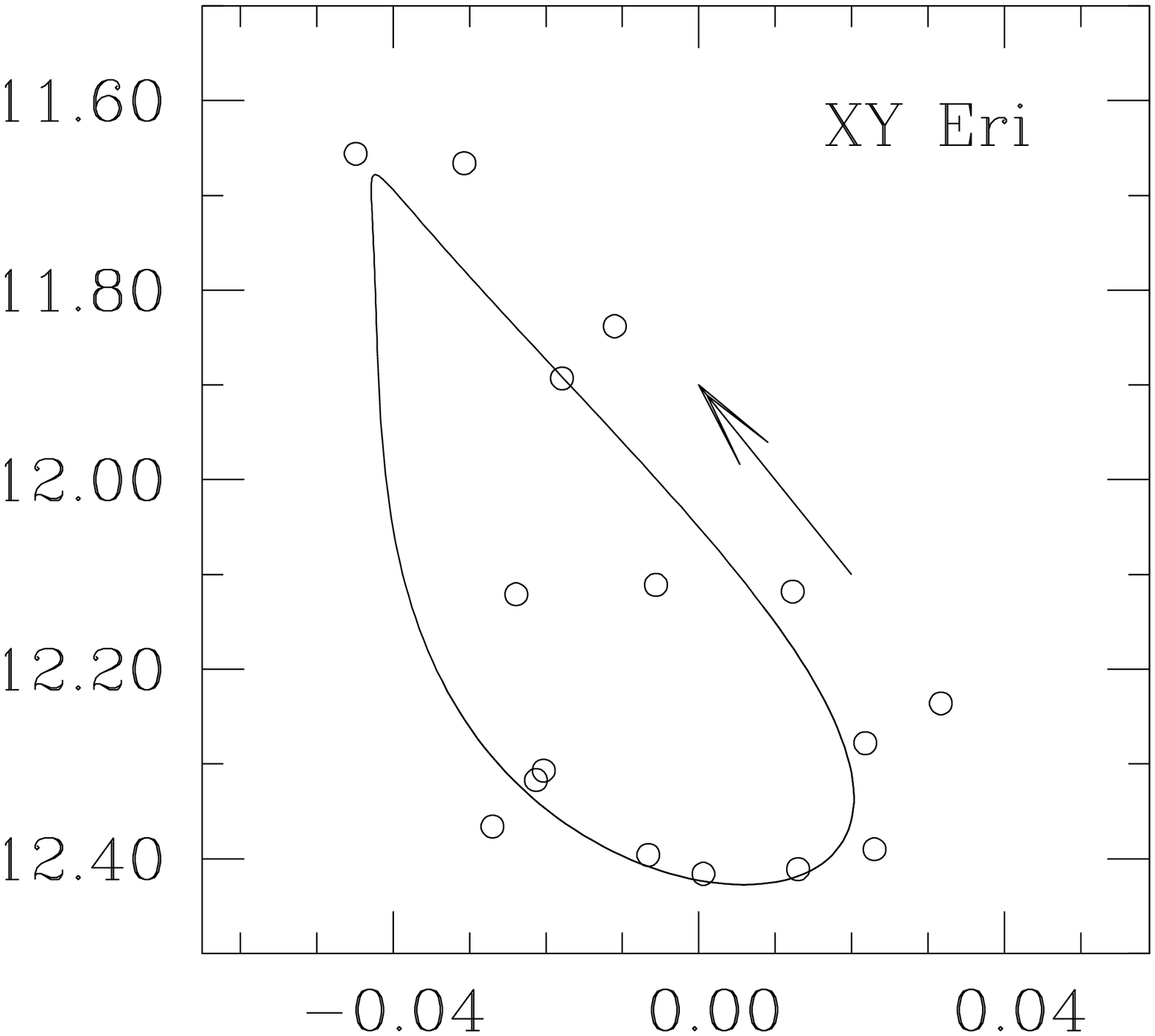}  &
   \includegraphics[width=4.9cm]{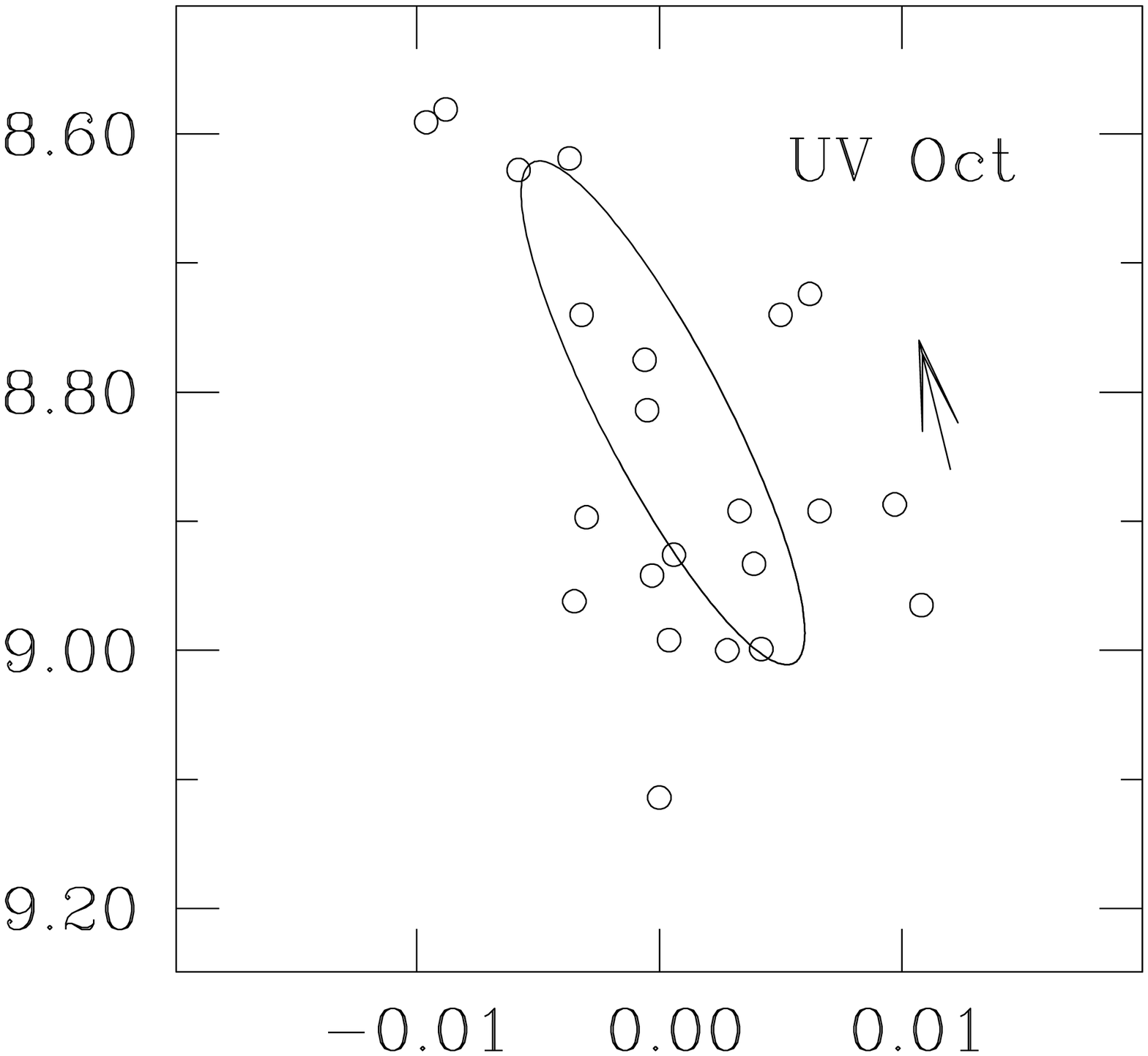}  &
   \includegraphics[width=4.9cm]{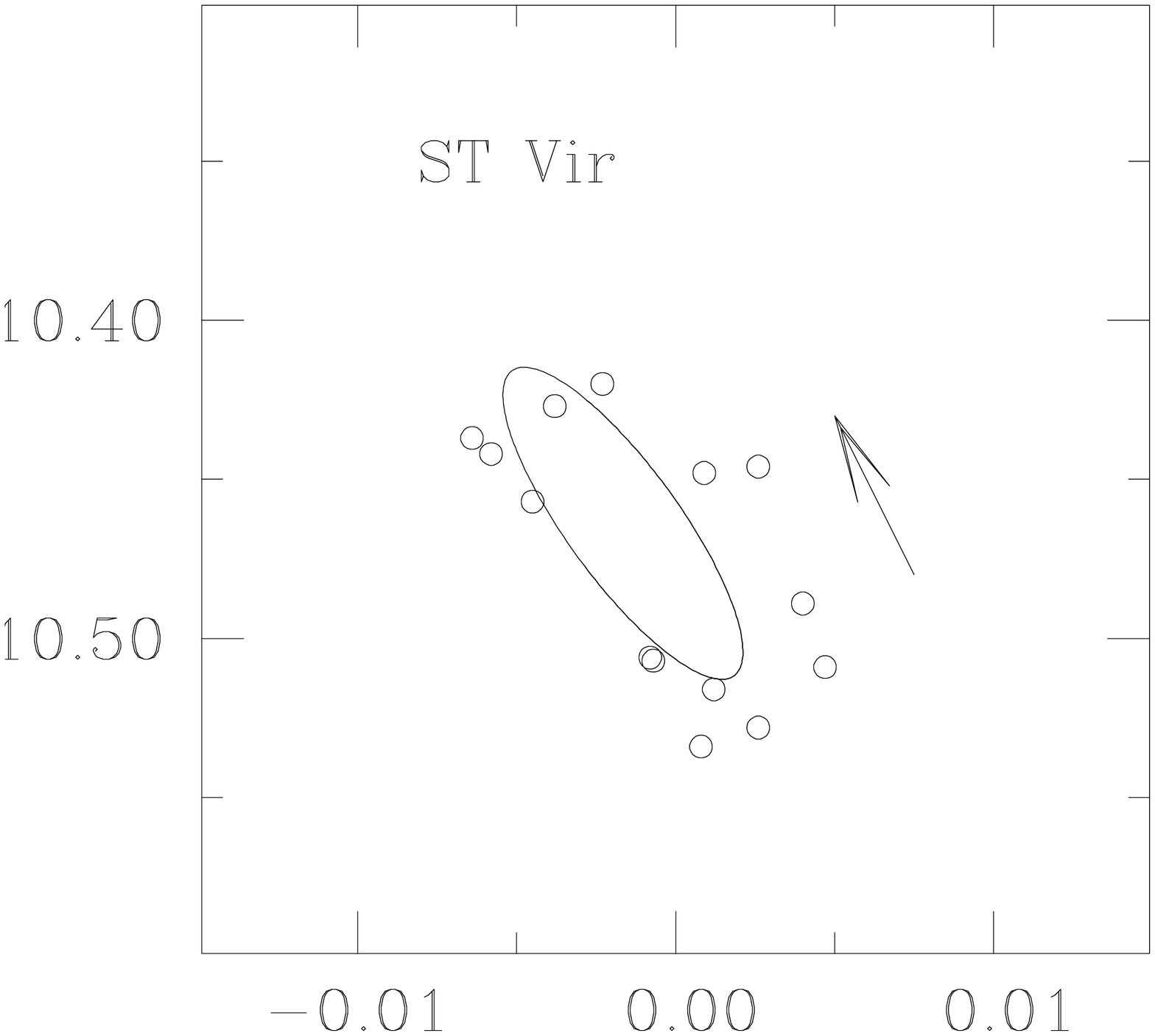}  \\
   \end{tabular}}
\caption{Stars with the brighest \vmax\, (ordinate) corresponding to the maximum negative O-C
(abscissa).  The closed curves are run in the anticlockwise direction.}
\label{potato3}
\end{figure*}
\begin{figure*}[ht]
\centerline{   \begin{tabular}{lcr}
   \includegraphics[width=4.9cm]{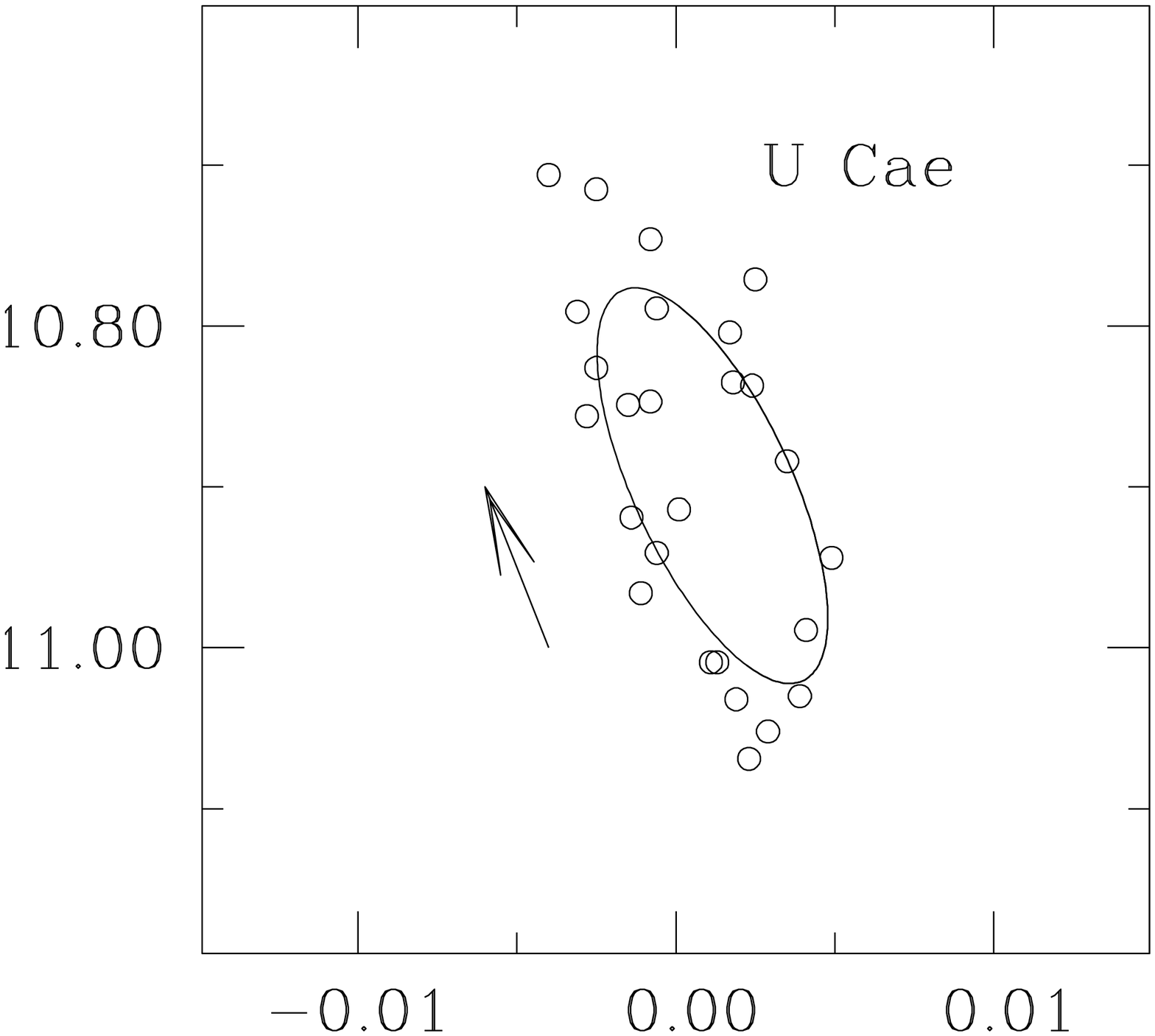}  &
   \includegraphics[width=4.9cm]{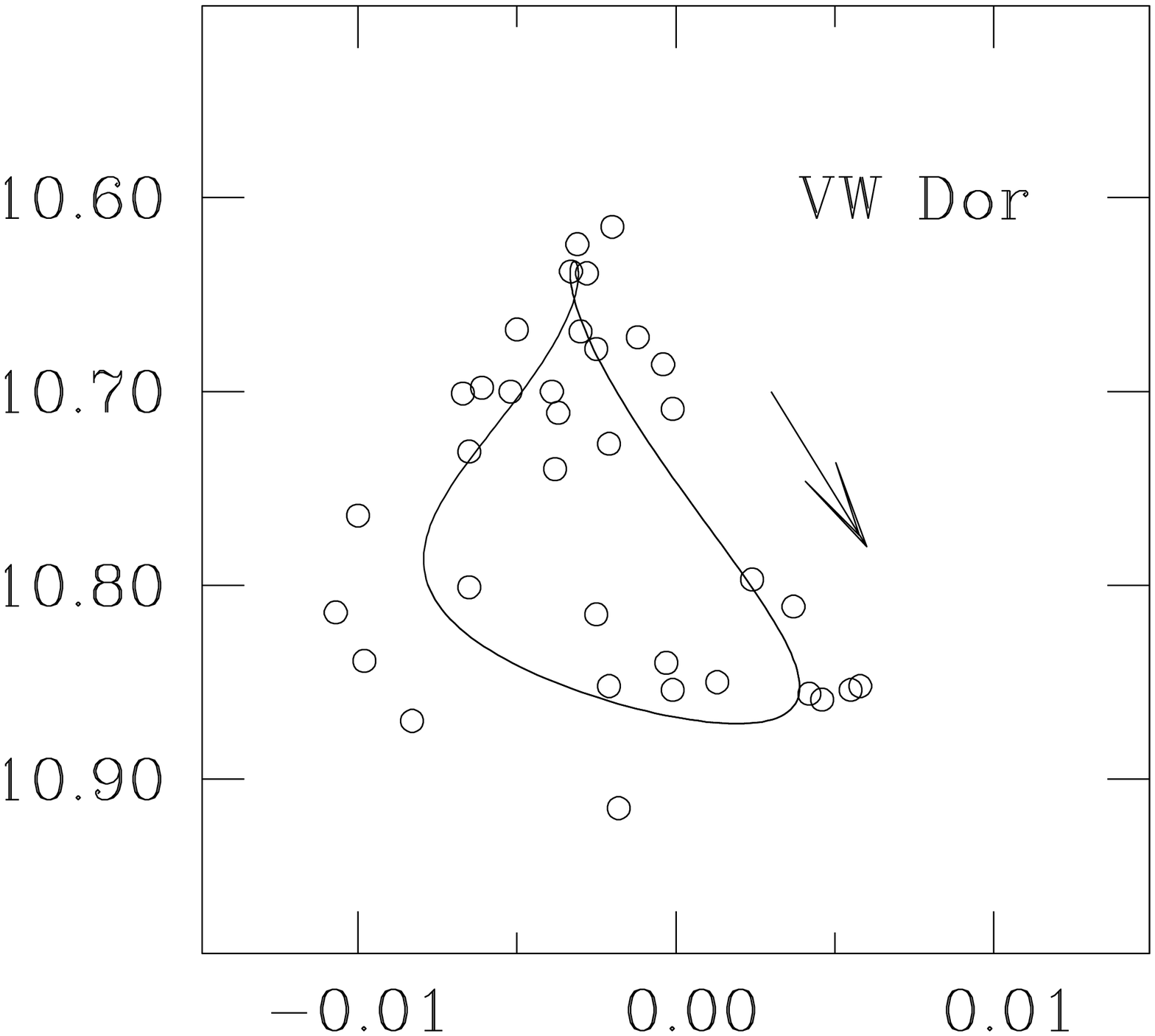}  &
   \includegraphics[width=4.9cm]{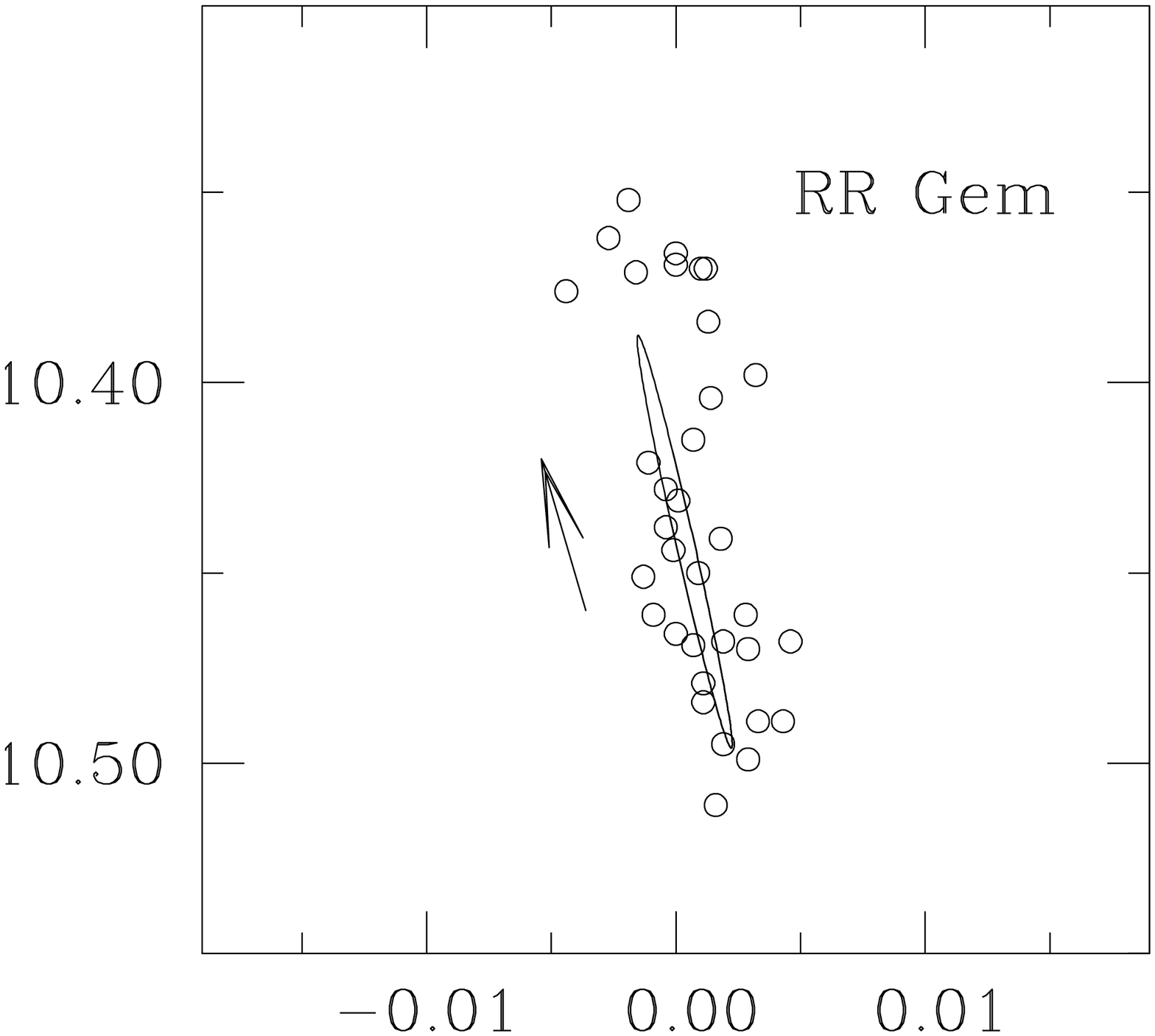}  \\
   \includegraphics[width=4.9cm]{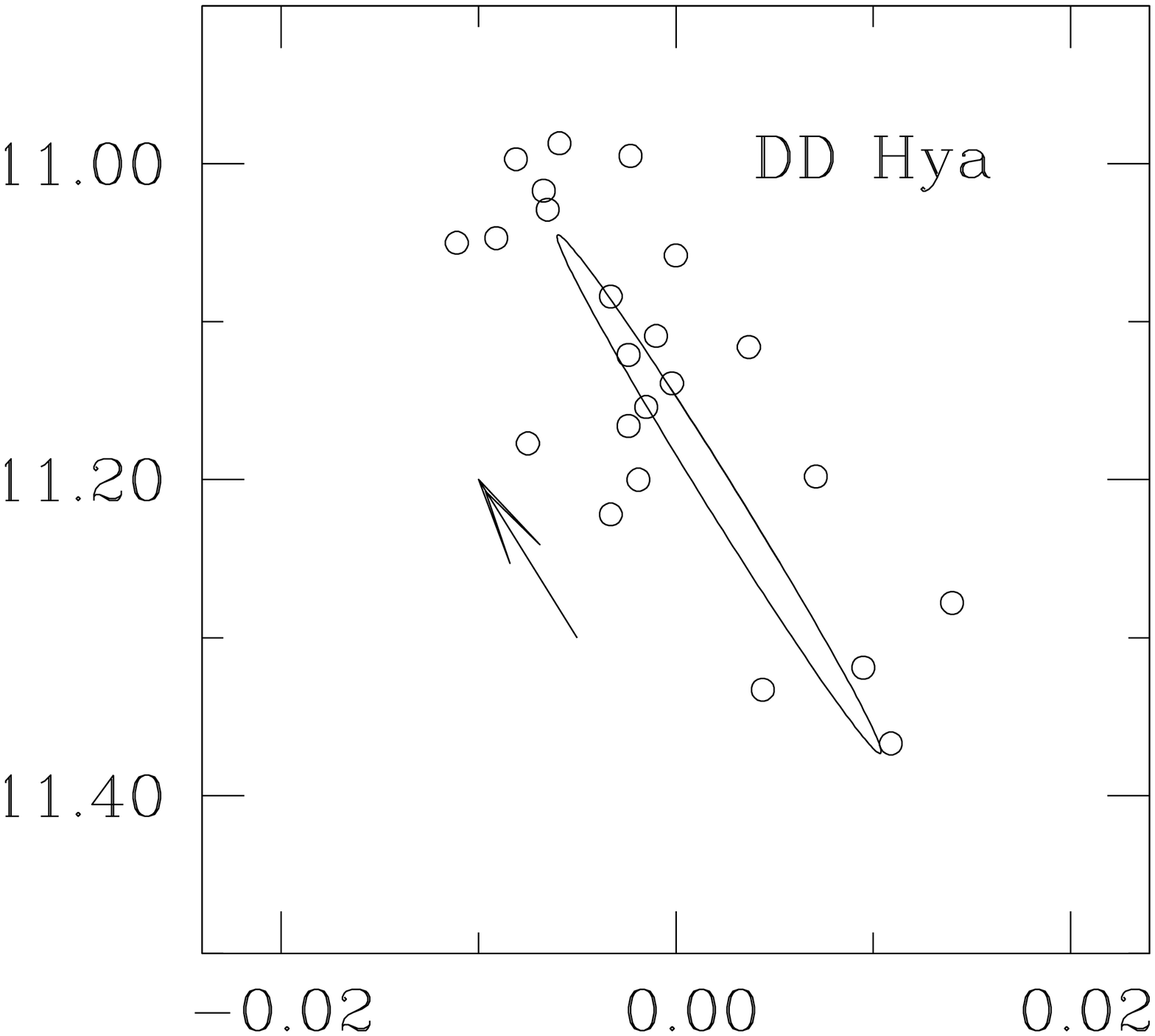}  &
   \includegraphics[width=4.9cm]{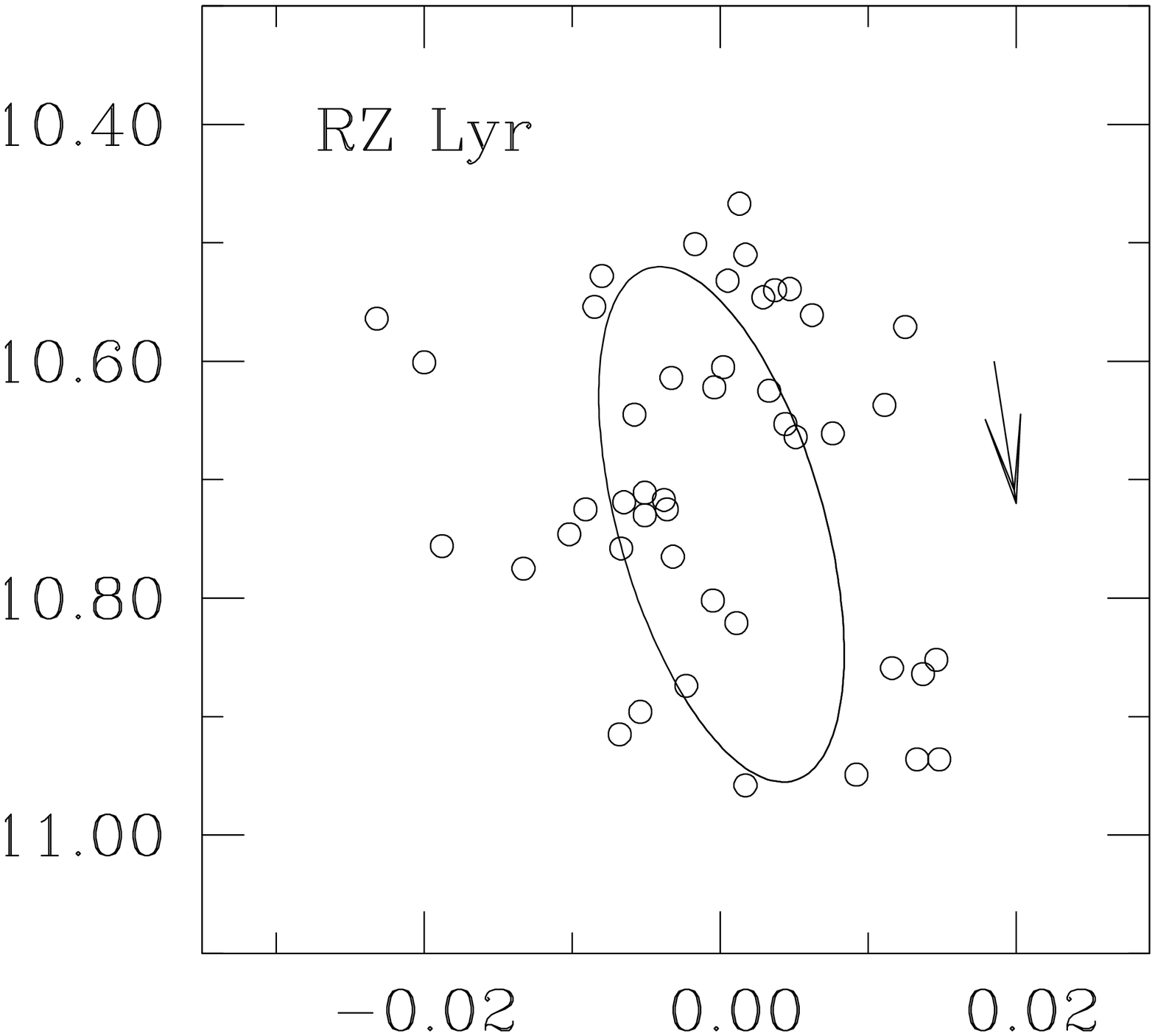}  &
   \includegraphics[width=4.9cm]{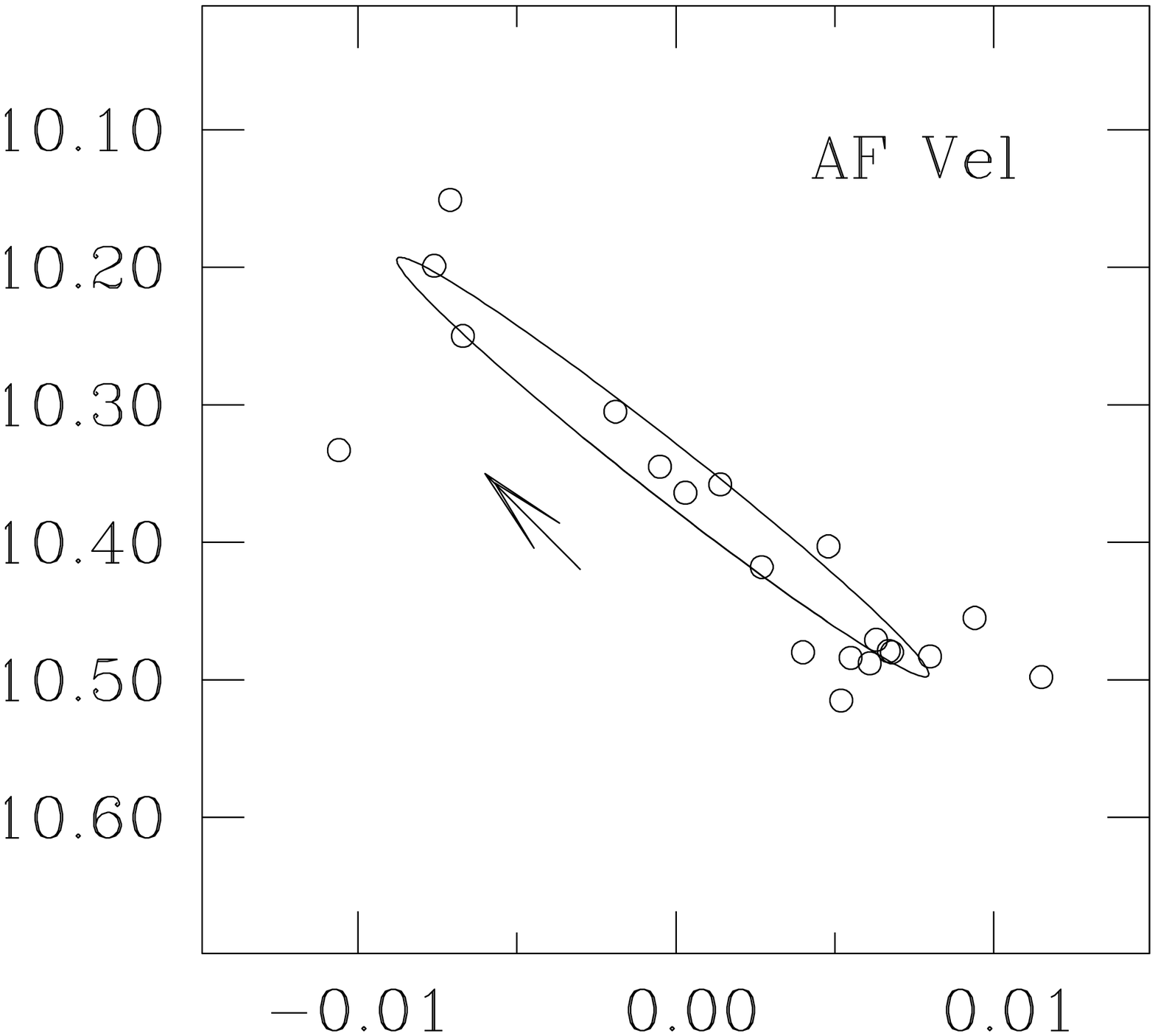}  \\
  &
   \includegraphics[width=4.9cm]{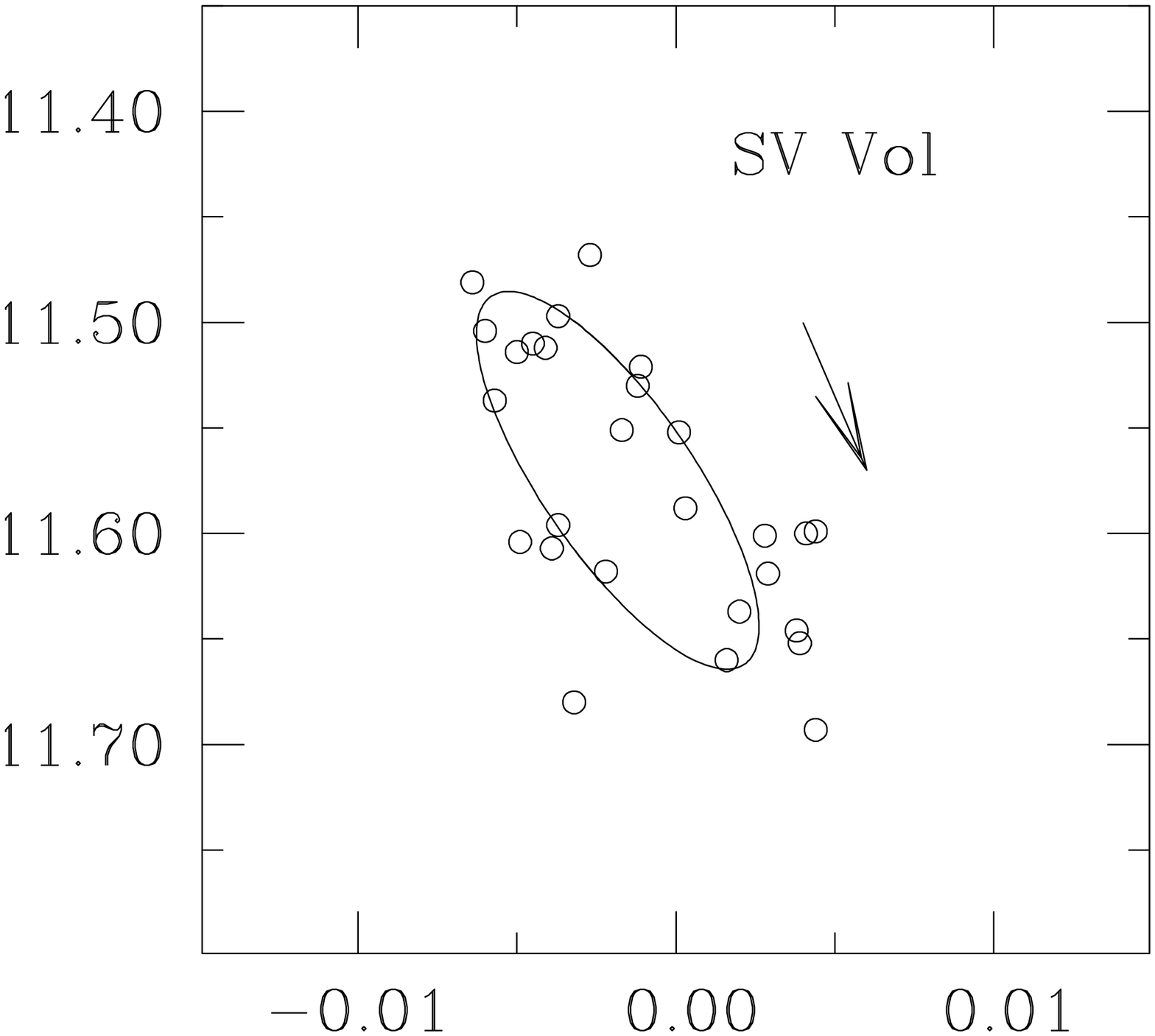}  &
  \\
   \end{tabular}}
\caption{Same as Fig.~\ref{potato3}, but the closed curves are run in the clockwise direction.}
\label{potato4}
\end{figure*}
The situation is quite different for the stars showing a
long Blazhko period, i.e., RX Col and RY Oct. At frequencies of 0.008~d$^{-1}$
(corresponding to a period of 130~d, as that of RX Col) or lower (RY Oct)
the alias at $-1$~y$^{-1}$  points to an almost twice longer period.
The extreme case is that of RU Cet.
The O-C's, the \vmax, and the folded light curve
suggest the possibility of a long period, probably much longer
than 100~d, but the strong aliasing effect prevented  us from
proposing a reliable value.
\\
In addition to those listed in Table~\ref{gen1},
other RRab stars in the GEOS database show cycle-to-cycle
variation which could be ascribed to a Blazhko effect, but
we could not propose a reliable value for the  period
due to the strong alias uncertainty.  The scientific discussion
of the Blazhko effect could be sharpened if new values for \tmax\, are collected on
these stars.
\section{Discussion}
\subsection{Varieties in the  phenomenology of the Blazhko effect}
With respect to the frequency analysis of ground- or space-based photometric
timeseries, the GEOS database enriched with the maxima observed with the TAROT
telescopes
offers the possibility of investigating  the relation between  amplitude
and phase changes,  described by  the O-C vs. \vmax\, diagram.
\citet{benko} recently interpreted the Blazhko light
curves as modulated signals and provided theoretical predictions of these
diagrams (see their Figs.~17 and 18).
\\
To exploit this possibility, we used the O-C and \vmax\, folded with the Blazhko period
(see bottom left panels in  Fig.~\ref{bddra} for an example). We calculated the least-squares
fits of these folded data, thus obtaining an O-C, \vmax\,  couple at any given Blazhko phase.
We drew the closed curve that describes the shape of the Blazhko effect by connecting
the consecutive couples.
A quick analysis immediately reveals how such closed curves  can assume
different shapes and can be run in opposite directions. We have 20 cases of anticlockwise
directions and 9 cases of clockwise ones.
The most numerous category is formed by Blazhko cycles where the
O-C and \vmax\, curves are roughly in phase and the closed curves are run in the
anticlockwise directions (13 cases, Fig.~\ref{potato1}). The $\Delta \phi$ shifts
range from $-0.06~P_B$ (RS Boo) to $-0.26~P_B$ (TT Cnc).
The curves of Z~CVn ($\Delta \phi=-0.79~P_B$) and UU~Hya ($-0.83~P_B$)
have the same  orientation, but run clockwise (2 cases, Fig.~\ref{potato2}).
\\
The O-C and \vmax\, curves are roughly in anti-phase and the closed curves are run
anticlockwise  for  six stars (Fig.~\ref{potato3}). The $\Delta \phi$ shifts range from
$-0.30~P_B$ (RX Col) to $-0.42~P_B$ (UV Oct). For this orientation of the closed curve the number of stars
running clockwise is nearly the same (seven cases, Fig.~\ref{potato4})  and
the phase shifts ranges from $-0.52~P_B$ (DD Hya) to  $-0.66~P_B$ (RZ Lyr).
\\
There  are close similarities with the curves predicted by the amplitude and frequency modulations
\citep{benko}.
In particular, the extreme cases of the Blazhko diagrams shown by BD Dra, SZ Hya
(Fig.~\ref{potato1}),  and VW~Dor (Fig.~\ref{potato4}) confirm
the complicated shapes predicted in case of simultaneous non-sinusoidal amplitude and phase modulations.
They are characterized by strongly elongated shapes, with some cuspids and loops,
due to the peculiar plots of the O-C and/or \vmax\, values. In the case of SZ Hya,
(Fig.~\ref{all} and Fig.\ref{potato1}) the $\Delta \phi$ shifts
reflect this  peculiarity and is different from the others of the same group of stars.
\\
These subdivisions reflect the mathematical differences in the signal modulations stressed by
\citet{benko}.
The phase shifts $\Delta \phi$
(Table~\ref{gen1}) can be tranformed in the phase parameter $\phi_m$ \citep[Eq.~45 and Fig.~17 in ][]{benko}
by the relation $\phi_m=\Delta \phi + k\,180$, where $k=1$ for the anticlockwise direction, $k=-1$ for
the clockwise direction.
It is noteworthy that \citet{benko} obtained their diagrams by a mathematical
representation of the light curve, while ours were built by using the observed
\tmax\, and \vmax\, values, without any assumption on the form of the
light curve. We can argue that our observational results are a rather solid confirmation that
the Blazhko effect can be  described by  amplitude and frequency modulations, though
the physical reasons of these modulations still remain unknown.
\subsection{Distribution of the Blazhko periods}
\begin{figure}
\epsscale{.80}
\centerline{   \includegraphics[width=8.5cm]{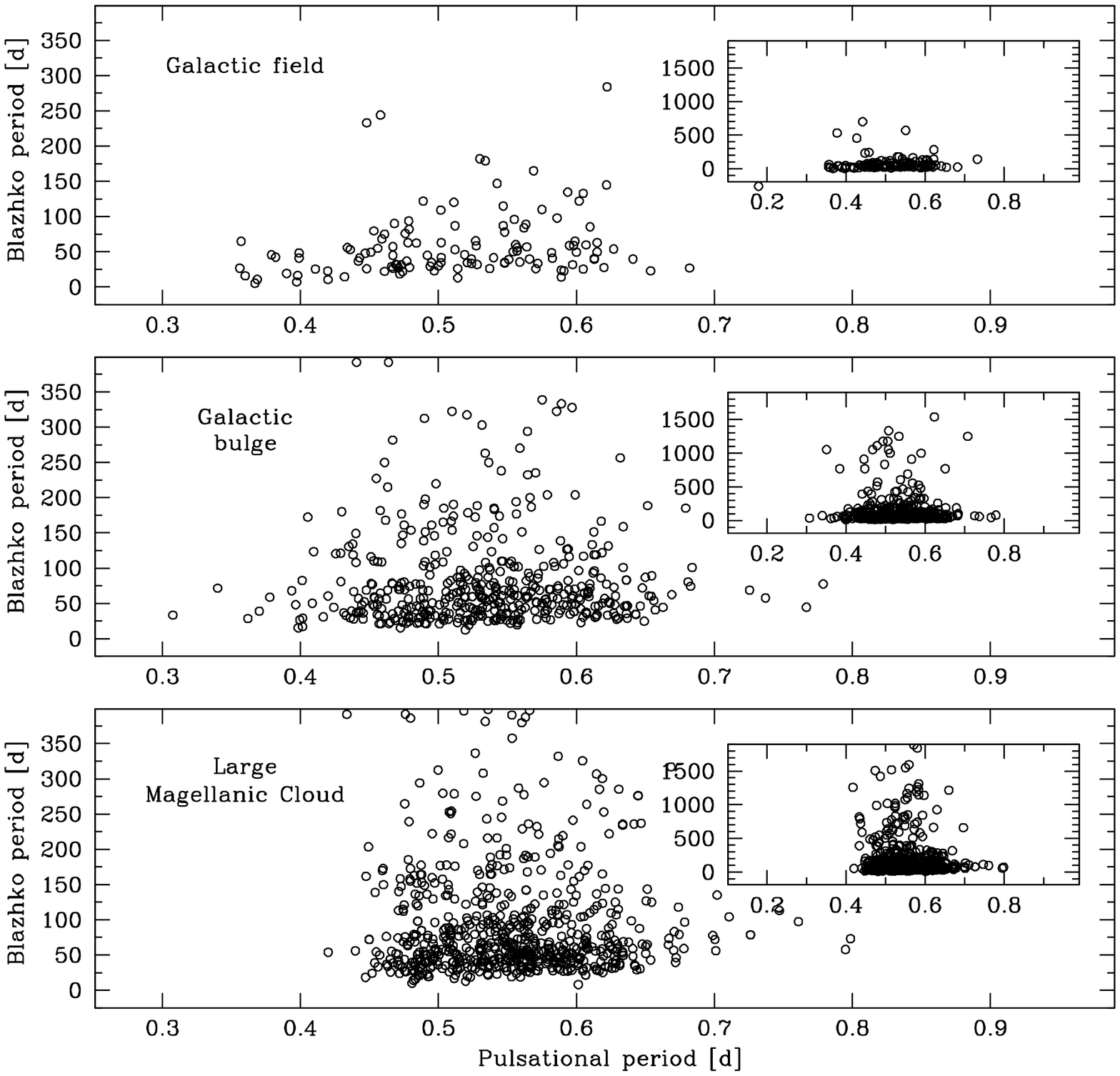} }
\caption{The  distribution of the Blazhko periods vs. the
pulsational periods for the sample in the
all-sky survey described in this paper ({\it top panel}),
in the OGLE galactic bulge ({\it middle panel})
 and in the MACHO LMC survey ({\it bottom panel}).
}
\label{distri}
\end{figure}

The characteristics of the all-sky galactic Blazhko RR Lyr stars could  be compared
with those obtained in the galactic bulge by OGLE \citep[][]{ogle2} and in
the Magellanic Clouds by MACHO \citep{macho}. The results of the OGLE and MACHO surveys were obtained
by performing a frequency analysis of the photometric timeseries and
the Blazhko periods were measured from the
separation of the multiplets in the amplitude spectra, without having the possibility
to determine individual \tmax\, and \vmax.
\\
We  analysed the OGLE stars considering the different subclasses introduced by
\citet{ogle2} on the basis of the appearance of their amplitude spectra:
BL1 (a single peak close to the main pulsational frequency),
BL2 (an equidistant triplet),
and BL2+? (equidistant triplet plus an additional peak) stars.
The distributions of the Blazhko periods vs. the pulsational ones are
not very different in the  plots of each subclass. The same applies for the similar
categories used in the MACHO classification. Figure~\ref{distri} allows a direct
comparison, although  the all-sky Blazhko variables are much less numerous than those
in the galactic bulge and in the Large Magellanic Cloud.
Blazhko periods longer than 400-500~d are not very common
(see inserted boxes).
The lack of such long periods in the all-sky survey could be  an
observational bias due to the impossibility of following an isolated star
along many years.
\\
The range of pulsational
periods is quite similar and the very few Blazhko variables below  0.40~d and above 0.67~d
reflects  the lesser number of RRab stars having these periods. The
long Blazhko periods for stars with $P<0.40$~d (RS Boo and of a couple of stars in
the galactic bulge)  are quite rare.
Note how the limited all-sky sample enhances an apparent linear increase
of the Blazhko period from 0.40 to 0.55~d, accompanied by a rather
flat distribution starting from 0.47~d toward longest pulsational periods. This
peculiarity is less discernible in the OGLE sample, but
disappears in the MACHO one.
We are not very  confident of the reality of such  structures. Indeed,
the histograms of the Blazhko periods did not reveal
any significant peak after the highest at 45~d in the MACHO sample and at 25~d in
the OGLE one.
\\
We prefer to consider Fig.~\ref{distri} as an evidence that the Blazhko period could
be very different for the same pulsational period (vertical scatter)
and that the same Blazhko period could be observed in a large
range of pulsational periods (horizontal scatter).
In such a context, the Blazhko effect of AH Cam (Sect.~\ref{ahc}) is very similar
to that of V1127 Aql \citep{aql} for the amplitudes of
the magnitudes and phases of the maxima, for the shape of the Blazhko cycle in the
O-C vs. magnitudes plot. The pulsational periods are very similar too. But the two
Blazhko periods are very different, although both are short ($P_B$=10.829~d for AH Cam,
$P_B$= 26.88~d for V1127 Aql).
\\
The analysis of the LMC stars in the MACHO sample allows  another
comparison. \citet{macho} report a higher
incidence (74\%) for frequency patterns in
which the modulation amplitude on the higher frequency sidepeaks
is larger than its lower frequency counterpart.
The mathematical formalism  \citep{benko} tells  us that
larger modulation on the higher frequency side indicates
anticlockwise direction in the \tmax\, vs. O-C plots.
As in the case of our galactic sample, this direction is
the preferred one for LMC RRab stars.
\section{Conclusions}
The GEOS database and the TAROT survey provided a treasure of information
on the Blazhko effect.
 TAROT light curves revealed that the non-repetitivity of the
shoulder on the rising branch is a common feature in both Blazhko and non-Blazhko stars.
This immediately translates into a model constraint, since the physical cause creating
the shoulder must work in both types of stars and the
Blazhko effect is not damped by this cause. The hydrodynamical models should take
into account  several features (the bump at minimum, the shoulder
on the rising branch, and also the double maximum observed in RRc stars) that accurate photometry
is putting in evidence on non-Blazhko stars, too. In this context, the new models explaining
the period doubling bifurcation \citep{szabo} should be used to test these features also.
\\
The  Blazhko effect appears to be multifaceted.
We observe very different
shapes of the O-C vs. \vmax\, plots and the described paths (circular, elongated, almost linear,
with loops and hooks,~...) can be run in a clockwise or anticlockwise direction also in
stars with similar periods, both pulsational and Blazhko.
Our observational results confirm the variety of shapes predicted by
a mathematical formalism involving amplitude and phase variations \citep{benko}.
As a further feedback to theoretical works,
the physical cause of the Blazhko mechanism must be able to explain both such a large variety
and  the preferred anticlockwise direction.
The very-long photometric monitoring also suggests that the Blazhko effect shows
cycle-to-cycle changes
\citep[also in the case of RR Lyr itself; ][ Le Borgne et al. in preparation]{preston}
and this further variability has still to be implemented in the formalism. We could say that
perfect regularity is the exception (i.e., BI Cen) rather than the
rule.
\\
We emphasize the occurrence of
different Blazhko periods at the same pulsational periods and of the same
Blazhko period at very different pulsational periods.
We could be tempted to consider that the
oblique-pulsator model combined with the randomly distributed line-of-sights and
(very) different rotational periods
could  produce the variety of the observed  O-C vs. \vmax\, plots. However, the non-detection
of relevant magnetic fields and the large cycle-to-cycle variations in the Blazhko effect
pose insoluble problems to this explanation.
\\
Since RRab stars are located in
a narrow strip on the horizontal branch, their physical conditions are not expected
to vary too much.  \citet{pervar} showed how
both redward and blueward shifts are observed in the evolutionary paths of RRab
stars and perhaps the evolutionary stage (i.e., contraction or expansion) could introduce
relevant differences in the Blazhko mechanism.
In the case of Z CVn we could verify
that the variation of the pulsational period resulted in changes
in the Blazhko effect, both in terms of period and amplitude,
and that they are anticorrelated.
Since the variations of the pulsational period reflect changes
in the star's radius, the Blazhko effect seems very sensitive to this
stellar parameter.  In this context, the unique case of correlated changes between
the Blazhko and pulsational periods (i.e., XZ~Dra), the few stars
with $P<0.40$~d showing  a very-long Blazhko period
and those showing  complicated Blazhko plots
can be considered promising laboratories in which to test future improvements
in the theoretical models of the Blazhko effect.
\acknowledgments
EP acknowledges Observatoire Midi-Pyr\'en\'ees for the three-month grant
allocated between May and July 2011 which allowed him to  work at the
{\it Institut de Recherche en Astrophysique et Plan\'etologie} in Toulouse, France.
The basic ideas  of this project have been sketched during several GEOS meetings,
where the different knowledge of amateur and professional astronomers found
a very profitable synthesis. The active participation of N.~Beltraminelli, M.~Benucci,
R.~Boninsegna, E.~Denoux, J.~Fabregat, F.~Fumagalli, D.~Husar, A.~Manna, J.C.~Misson,
P.~de~Ponthiere, J.~Remis, and J.~Vialle to these discussions is gratefully acknowledged.
The participations of M.~Correa, F.~Salado, D.~Fern\'andez, and R. Moliner
to the observations at the {\it Agrupaci\'o Astron\`omica de Sabadell} Observatory
and of T.~Krajci to the operations of the  AAVSOnet telescopes are
gratefully acknowledged.
The present study has used the SIMBAD database operated at the {\it Centre de
Donn\'ees Astronomiques} (Strasbourg, France)  and the GEOS RR Lyr database
hosted by IRAP (OMP-UPS, Toulouse, France).

\appendix
\section{The inventory of galactic Blazhko stars}
The inventory of Blazhko stars located in the galactic field has been proposed and
revised several times. The first comprehensive list was published by \citet{sze1,sze2}
and then adapted by \citet{smith} for his book. \citet{lloyd} discovered new Blazhko
RR Lyr stars by analyzing the {\it Northern Sky Variability Survey} \citep[NSVS, ][]{nsvs}.
\citet{sodor} did the same by analyzing the
{\it All Sky Automated Survey} \citep[ASAS, ][]{asas}.
\citet{rev1} corrected some uncertain results and \citet{rev2} added some
complementary stars, partially
overlapping  those proposed by \citet{lloyd}.
The Blazhko effect was discovered on many RRab stars located in the {\it Kepler} field-of-view
\citep{kepler}.  Table~\ref{gen2} lists the  Blazhko RRab stars complementing Table~\ref{gen1}.
New entries are expected from the KBS~II, e.g., V397 Her and ASAS~212034+1837.2 \citep{granada},
the Sloan Digital Sky Survey \citep{sdss},
and from other specific projects, as SATINO\footnote{http://stiftung-astronomie.de/satino.htm}
and FARO \citep{faro}.
\\
Table~\ref{gen2} does not include the Blazhko stars in the galactic bulge discovered in the OGLE project,
for which we refer the reader to the classification paper \citep{ogle2}. Moreover,
the stars listed by \citet{lloyd}, \citet{budapest}, and \citet{kepler}
with an uncertain Blazhko period are not considered as well.
This is justified by the  59:~d period suggested for RZ Lyr,
while the true period is 120~d \citep{rzlyr}. Other cases are controversial:
FM Per (20~d after \citealt{lloyd}, 122~d after \citealt{lee}), AR Ser (63~d  after \citealt{lloyd}, 110~d
after \citealt{lee}), and X Ret (45~d after \citealt{smith}, 161~d after \citealt{sodor}).
These uncertainties are not surprising since sometimes it is very difficult to have a clear evidence:
the Blazhko effect of DM Cyg was announced  by \citet{lysova},
then  questioned by \citet{rev1}, and finally  confirmed by new observations \citep{dmcyg}.
\\
Except for the stars too faint to be measured with 25-cm telescopes,
we are planning to complete their monitoring in the next years.
In such a way we could extend the TAROT
homogenous sample of well--studied Blazhko stars.
\begin{deluxetable*}{ lll cl c lll  cl}
\tablecolumns{10}
\tablewidth{0pc}
\tabletypesize{\footnotesize}
\tablecaption{
RRab stars  showing a known Blazhko effect and not discussed in this paper.}
\tablehead{
\multicolumn{2}{c}{Star} &
\multicolumn{1}{c}{$P_{\rm puls}$} &
\multicolumn{1}{c}{Known $P_B$} &
\multicolumn{1}{c}{Reference} &&
\multicolumn{2}{c}{Star} &
\multicolumn{1}{c}{$P_{\rm puls}$} &
\multicolumn{1}{c}{Known $P_B$} &
\multicolumn{1}{c}{Reference} \\
\multicolumn{2}{c}{} &
\multicolumn{1}{c}{[d]} &
\multicolumn{1}{c}{[d]} &
\multicolumn{1}{c}{} &&
\multicolumn{2}{c}{} &
\multicolumn{1}{c}{[d]} &
\multicolumn{1}{c}{[d]} &
\multicolumn{1}{c}{} \\
}
\startdata
SW    & And        & 0.442 &  36.8  & (1)     && V445  & Lyr        & 0.512 & 53.1   & (9),(10)\\
XY    & And        & 0.399 &  41.2  & (6)     && V450  & Lyr        & 0.504 & 125:   & (9)     \\
DR    & And        & 0.564 &  57    & (5)     && RS    & Oct        & 0.458 & 244    & (2)     \\
GV    & And        & 0.528 &  32    & (4)     && DZ    & Oct        & 0.477 &  36.8  & (2)     \\
OV    & And        & 0.471 &  27    & (5)     && V788  & Oph        & 0.547 & 115    & (1),(3) \\
V1127 & Aql        & 0.356 &  26.86 & (7)     && V829  & Oph        & 0.569 & 165    & (1),(3) \\
S     & Ara        & 0.451 &  49.5  & (2)     && V1280 & Ori        & 0.479 &  28    & (5)     \\
SW    & Boo        & 0.514 &  13    & (1),(3) && FO    & Pav        & 0.551 & 571    & (2)     \\
RW    & Cnc        & 0.547 &  87    & (1)     && AE    & Peg        & 0.497 &  23    & (5)     \\
SS    & Cnc        & 0.367 &   5.3  & (6)     && BH    & Peg        & 0.641 &  39.8  & (1)     \\
RV    & Cap        & 0.448 & 233    & (1)     && FM    & Per        & 0.489 & 122 20 & (8),(5) \\
V674  & Cen        & 0.494 &  29.5  & (1)     && CS    & Phe        & 0.484 &  62.5  & (2)     \\
RY    & Com        & 0.469 &  32    & (6)     && AL    & Pic        & 0.548 &  34    & (2)     \\
TU    & Com        & 0.461 &  75    & (1),(3) && SW    & Psc        & 0.521 &  34.5  & (1),(3) \\
WW    & CrA        & 0.559 &  35.5  & (2)     && X     & Ret        & 0.492 &  45    & (1)     \\
X     & Crt        & 0.732 & 143    & (2)     && BT    & Sco        & 0.548 &  78    & (2)     \\
XZ    & Cyg        & 0.467 &  57.3  & (1)     && V494  & Sco        & 0.427 & 455    & (2)     \\
DM    & Cyg        & 0.420 &  10.57 & (6)     && AR    & Ser        & 0.575 & 110 63 & (1),(5) \\
V759  & Cyg        & 0.360 &  16.0  & (6)     && BR    & Tau        & 0.390 &  19.3  & (6)     \\
V783  & Cyg        & 0.620 &  27.7  & (9)     && UX    & Tri        & 0.467 &  45    & (4),(5) \\
V808  & Cyg        & 0.547 &  90:   & (9)     && UZ    & UMa        & 0.467 &  26.7  & (6)     \\
V1104 & Cyg        & 0.436 &  53.1  & (9)     && AD    & UMa        & 0.548 &  35    & (1),(3) \\
V2178 & Cyg        & 0.486 &  $>$200  & (9)     && NS    & UMa        & 0.599 &  65    & (5)     \\
WY    & Dra        & 0.589 &  14.3  & (1),(3) && UZ    & Vir        & 0.459 &  68.2  & (6)     \\
XZ    & Dra        & 0.476 &  76    & (1)     && AM    & Vir        & 0.615 &  49.8  & (2)     \\
AV    & Dra        & 0.555 &  96    & (5)     && FK    & Vul        & 0.434 &  56    & (6)     \\
SS    & For        & 0.495 &  34.9  & (2)     &&       &            &       &        &         \\
RT    & Gru        & 0.512 &  87    & (2)     && CoRoT & 0101128793 & 0.472 &  18.66 & (7)     \\
AR    & Her        & 0.470 &  31.6  & (1),(5) && CoRoT & 0100881648 & 0.607 &  59.8  & (7)     \\
BD    & Her        & 0.474 &  22    & (6)     && CoRoT & 0101503544 & 0.605 &  25.6  & (7)     \\
DL    & Her        & 0.572 &  33.6  & (1),(5) &&       &            &       &        &         \\
V365  & Her        & 0.613 &  40.6  & (1),(5) && KIC   & 11125706   & 0.613 &  39.4  & (9)     \\
V421  & Her        & 0.557 &  56    & (5)     &&       &            &       &        &         \\
V434  & Her        & 0.514 &  26.1  & (1),(3) && GSC   & 0275-0090  & 0.595 &  59    & (5)     \\
V442  & Her        & 0.442 & 700    & (4)     && GSC   & 0318-0905  & 0.447 &  48    & (5)     \\
V1124 & Her        & 0.551 &  39    & (4)     && GSC   & 0607-0591  & 0.456 &  55    & (5)     \\
SV    & Hya        & 0.478 &  63    & (2)     && GSC   & 1581-1784  & 0.591 &  23    & (5)     \\
CZ    & Lac        & 0.432 &  14.6  & (6)     && GSC   & 1667-1182  & 0.562 &  84    & (5)     \\
SZ    & Leo        & 0.534 & 179    & (2)     && GSC   & 1948-1733  & 0.502 &  42    & (5)     \\
AH    & Leo        & 0.466 &  29    & (4)     && GSC   & 4378-1934  & 0.519 &  46    & (5)     \\
Y     & LMi        & 0.524 &  33.4  & (1)     && GSC   & 5590-0758  & 0.540 &  41.7  & (2)     \\
FU    & Lup        & 0.382 &  42.4  & (2)     && GSC   & 5828-0847  & 0.627 &  54    & (2)     \\
PQ    & Lup        & 0.582 &  48.8  & (2)     && GSC   & 5885-0757  & 0.602 & 122    & (2)     \\
RR    & Lyr        & 0.566 &  39.6  & (1),(9) && GSC   & 6619-1146  & 0.598 &  59    & (2)     \\
AQ    & Lyr        & 0.357 &  64.9  & (6)     && GSC   & 6672-0596  & 0.399 &  48.3  & (2)     \\
KM    & Lyr        & 0.500 &  30    & (1)     && GSC   & 6730-0109  & 0.448 &  26    & (4),(2) \\
MW    & Lyr        & 0.398 &  16.5  & (6)     && GSC   & 6811-0414  & 0.461 &  22.2  & (2)     \\
NR    & Lyr        & 0.682 &  27    & (5)     && GSC   & 6964-0926  & 0.530 & 182    & (2)     \\
V349  & Lyr        & 0.507 & $>>$127  & (9)     && GSC   & 7448-0418  & 0.379 &  45.7  & (2)     \\
V353  & Lyr        & 0.556 &  60.0  & (9)     && GSC   & 8297-1427  & 0.600 &  49.5  & (2)     \\
V354  & Lyr        & 0.561 & $>>$127  & (9)     && GSC   & 8814-0696  & 0.605 & 133    & (2)     \\
V355  & Lyr        & 0.473 &  31.4  & (9)     && GSC   & 8826-0640  & 0.615 &  63    & (2)     \\
V360  & Lyr        & 0.557 &  51.4  & (9)     &&       &            &       &        &         \\
V366  & Lyr        & 0.527 &  65.6  & (9)     && NSV   & 5200       & 0.502 &  63    & (2)     \\
\enddata
\tablecomments{See \citet{ogle2} for the OGLE stars in the galactic bulge.
References: (1)~\citet{smith}; (2)~\citet{sodor}, (3)~\citet{rev1},
(4)~\citet{rev2}, (5)~\citet{lloyd}, (6)~\citet{budapest}, (7)~\citet{santafe},
(8)~\citet{lee}, (9)~\citet{kepler}, (10)~\citet{V445}.
New names are provided by the {\it General Catalogue of Variable Stars} for the newly
discovered RRab Blazhko variables V1124~Her$\equiv$NSV~8170 \citep{rev2},
PQ~Lup$\equiv$NSV~7330, DZ~Oct$\equiv$NSV~4350, CS~Phe$\equiv$NSV~420, and
AL~Pic$\equiv$NSV~1700 \citep{sodor},
V1280~Ori$\equiv$NSV~2724, and NS~UMa$\equiv$NSV~4034 \citep{lloyd}.
}
\label{gen2}
\end{deluxetable*}
\end{document}